\documentstyle[emulateapj,apjfonts]{article}

\newcommand{\plotfour}[4]{\vbox{
\hbox{\epsfxsize=.45\columnwidth \epsfbox{#1} \hfil
\epsfxsize=.45\columnwidth \epsfbox{#2}}
\hbox{\epsfxsize=.45\columnwidth \epsfbox{#3} \hfil
\epsfxsize=.45\columnwidth \epsfbox{#4}}}}

\slugcomment{July 27th 1998}

\lefthead{Worrall et al.}
\righthead{The extreme AGN J2310-437}

\begin{document}
\submitted{Printed July 27th 1998. To appear in The Astrophysical Journal}

\title{A Multiwavelength Study of the Extreme AGN J2310-437}

\author{D.M.~Worrall\altaffilmark{1,2},
            M.~Birkinshaw\altaffilmark{1,2}, 
            R.~A.~Remillard\altaffilmark{3}, 
            A.~Prestwich\altaffilmark{2}, 
            W.~H. Tucker\altaffilmark{2}, 
            H.~Tananbaum\altaffilmark{2}, 
}
\altaffiltext{1}{Department of Physics, University of Bristol, U.K.}
\altaffiltext{2}{Harvard-Smithsonian Center for Astrophysics, Cambridge, MA 02138}
\altaffiltext{3}{Massachusetts Institute of Technology, Cambridge, MA
02139}

\authoremail{d.worrall@bristol.ac.uk}

\begin{abstract}

We present new X-ray, radio, and optical data for the unusual
galaxy/cluster system J2310-437. Our results confirm the presence of
an active nucleus, and suggest an interpretation as an anomalous
BL~Lac object of bulk relativistic Doppler factor $< 2$, with an
optically deficient radio-to-X-ray spectrum.  The radio, optical, and
soft X-ray flux densities could lie along a single power-law function,
lacking the curvature typical of BL Lac objects.  Compared with other
known sources that may have comparable multifrequency spectra,
J2310-437 is the most extreme.  Its low isotropic optical/UV radiation
is consistent with the intensity of external photons governing the
electron spectral break through Compton cooling; in this source the
external photon density would be too low to produce a spectral break
below the X-ray.

\end{abstract}

\section{Introduction}

The galaxy/cluster system J2310-437 is unusual.  CCD images and
spectroscopy identified the {\it Einstein\/} X-ray source E2307-44
(B1950 coordinates) with a galaxy of $V = 16.05 \pm 0.02$~mag and $z =
0.0886$ at the center of a cluster of Abell richness class~0
(\markcite{tuck95}Tucker, Tananbaum \& Remillard 1995).  ROSAT PSPC
observations showed X-ray extent (\markcite{tan97}Tananbaum et
al.~1997).  However, while the 0.5-2.5~keV luminosity\footnote{We use
$H_o = 50$~km s$^{-1}$ Mpc$^{-1}$, $q_o = 0$ throughout.  At the
redshift of J2310-437, 1 arcsec corresponds to 2.27 kpc.} of $\sim 7.5
\times 10^{43}$ ergs s$^{-1}$ would make J2310-437 one of the most
luminous known cluster X-ray sources for its richness class
({\it cf\/} \markcite{briel93}Briel \& Henry 1993), a plausible but non-unique
interpretation of the PSPC data was a combination of cluster and AGN
emission.  \markcite{tan97}Tananbaum et al.~(1997) favored associating
about 80\% of the X-rays with an AGN, noting the $\sim 20''$
positional coincidence of the galaxy with the 61~mJy radio source
J2310-437 from the PMN 4.85~GHz survey (\markcite{griff93}Griffith \&
Wright 1993; \markcite{cond93}Condon, Griffith \& Wright 1993).
However, they pointed out the lack of evidence for an AGN in the
optical, where the spectrum is typical of a normal elliptical galaxy,
devoid of emission lines.  The radio and X-ray strength combined with
the upper limit on optical continuum from the putative AGN component
would place it as an extreme BL~Lac object.

In this paper we present ROSAT HRI observations which verify the
compact nature of much of the X-ray emission.  Radio imaging and
polarization measurements with the Australia Telescope Compact Array
(ATCA) confirm the identification of the X-ray emitting galaxy with
the radio source.  We present a new optical spectrum from which we
derive improved upper limits to an optical AGN component.  Finally, we
discuss the classification of this extreme source, its relation to
other AGN populations, and the process that might be responsible for
its multiwavelength properties.

\section{ROSAT HRI X-ray Observations}

If indeed $\sim 80\%$ of the X-rays detected with the PSPC were from
an AGN, the better spatial resolution of the HRI was expected to
reveal a core of bright unresolved emission in our 50~ks exposure.
The ROSAT scheduling algorithm caused the HRI observations to be
carried out in two unequal observing periods roughly 6 months apart;
details are given in Table~\ref{hriobs}.
Our analysis was performed with the IRAF/PROS. Spatial
analysis also used generalized software (\markcite{birk94}Birkinshaw
1994; \markcite{worr94}Worrall \& Birkinshaw 1994) for fitting radial
profiles to combinations of models convolved with instrument Point
Response Functions (PRFs).

\begin{deluxetable}{lcrc}
\tablefontsize{\footnotesize}
\tablewidth{0pt}
\tablecaption{ROSAT HRI observations of J2310-437.
\label{hriobs}}
\tablehead{
\colhead{Dates}     & \colhead{Number}  &
\colhead{Total} & \colhead{Nominal} \\
& \colhead{of intervals}  &
\colhead{Livetime (ks)} & \colhead{Roll Angle}
}
\startdata
Nov 22 - 30, 1995 & 6 & 7.437 & $36^\circ$ \nl
May 1 - Jun 4, 1996 & 23 & 31.435 & $-141^\circ$ \nl
\enddata
\end{deluxetable}

\begin{figure*}
\plottwo{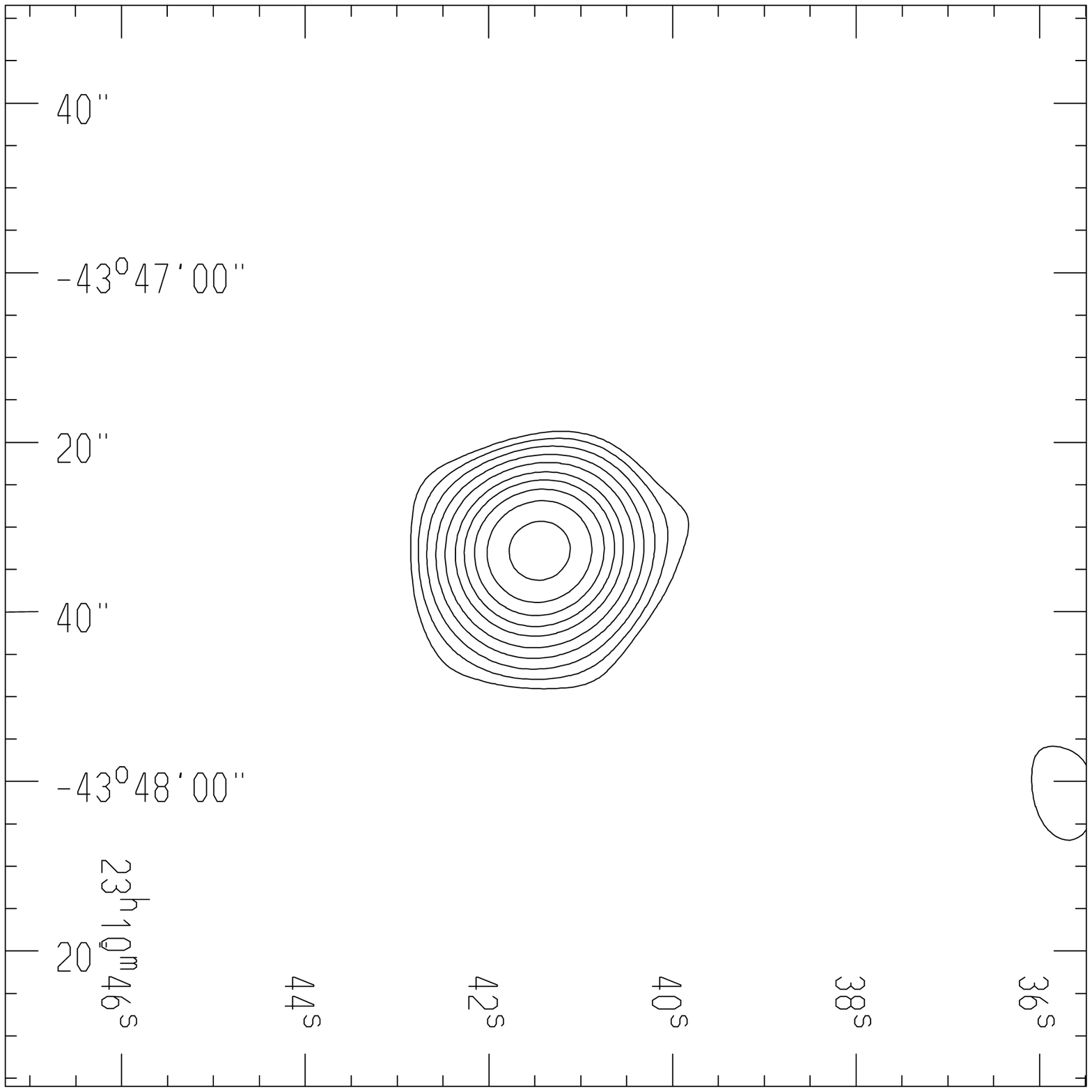}{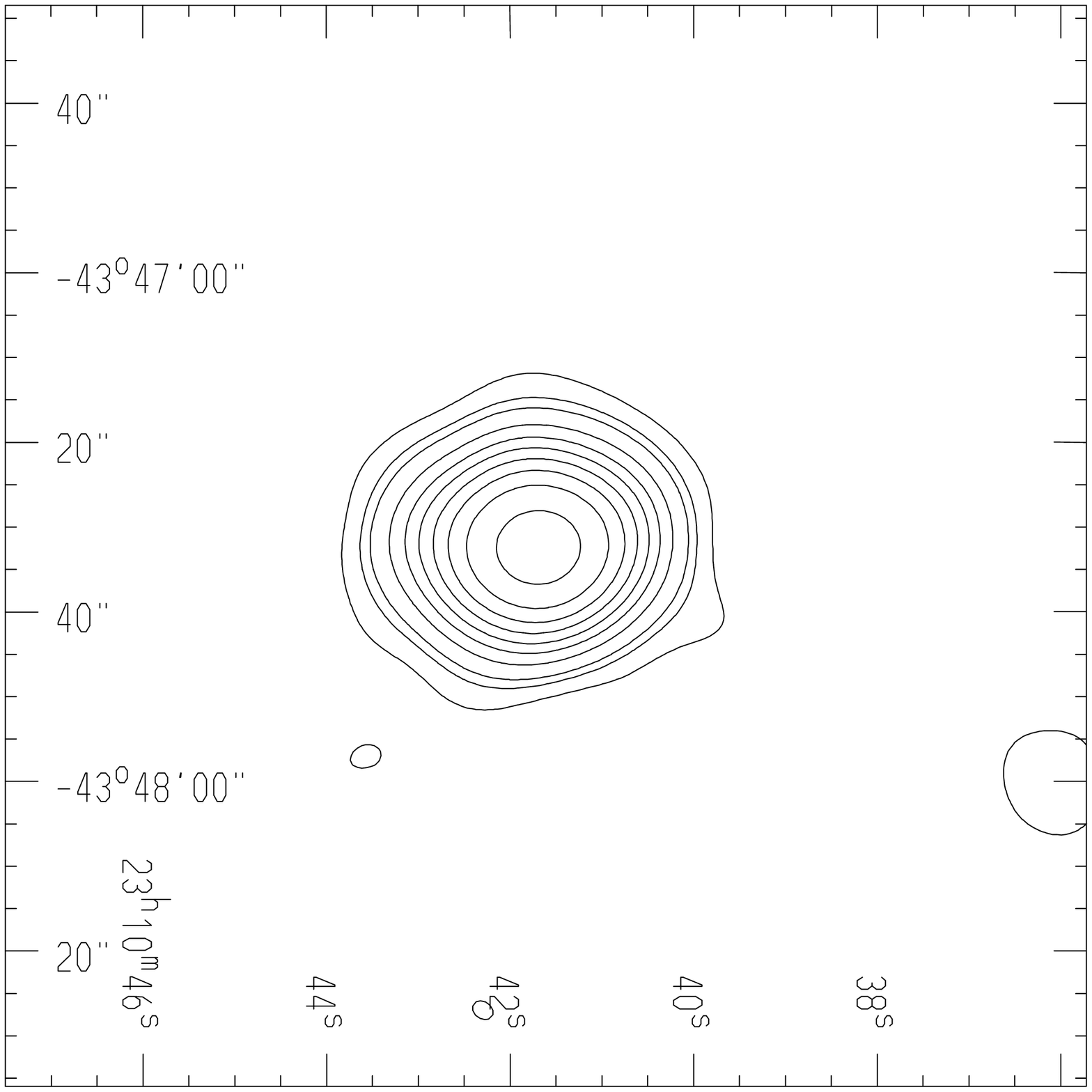}
\caption{HRI data from 1995 (left) and 1996 (right). Pixels are
$0.5 \times 0.5$ arcsec square and the plots show a $256 \times
256$ pixel region centered on J2310-427. The data have been smoothed
with a Gaussian of $\sigma = 4$ arcsec.  In each case the lowest
contour corresponds to a value which should occur only in four pixels
of the map by chance (i.e., $3.8\sigma$ significance), as determined
using a method appropriate for Poisson statistics 
(Hardcastle et al.~1998b).  For the 1995 data (which fit a point
source: Fig.~\ref{hri1profile}) the maximum is 0.91 cts/pixel
and contours are at 0.011, 0.018, 0.03, 0.05, 0.08, 0.13, 0.19, 0.28,
0.42, and 0.7.  For the 1996 data the maximum is 2.14 cts/pixel and
contours are at 0.018, 0.04, 0.06, 0.12, 0.2, 0.3, 0.44, 0.64, 0.92,
and 1.6.
\label{hriconts}}
\end{figure*}

The ROSAT HRI data for the region in the vicinity of J2310-437 are
shown smoothed and contoured in Figure~\ref{hriconts}.  The data from
1995 give radially symmetric contours whereas the 1996
data show clear ellipticity with the major axis running roughly
east-west.  The two data sets were taken at different spacecraft roll
angles, and the apparent difference in source
shape is undoubtedly because the 1996 data are more affected by residual
errors in the aspect corrections applied to the photons (Appendix
\ref{appendix}).  The impact on our analysis is that we use only the
(higher-quality) 1995 data to probe the spatial distribution of X-rays
associated with J2310-437; unfortunately these data represent just
less than 20\% of the total HRI exposure.  Both data sets have been
used to search for X-ray variability.

\subsection{Spatial Decomposition} \label{xrayspatial}

\begin{figure*}
\plotone{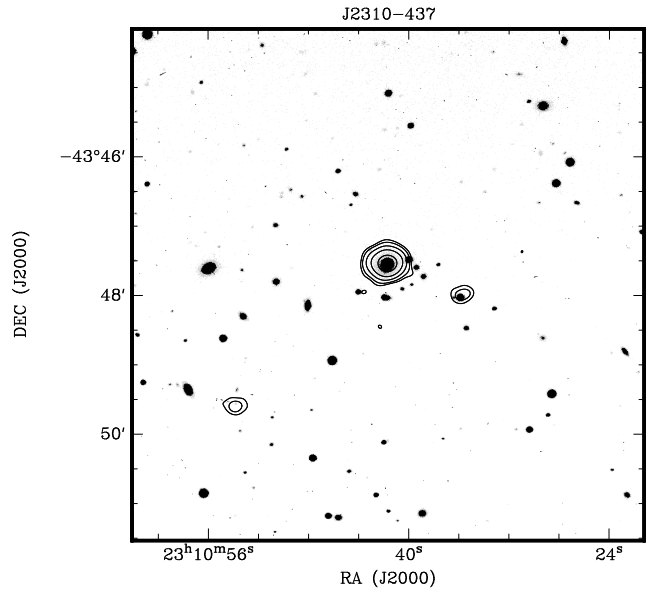}
\caption{HRI X-ray contours from the (longer) 1996 observation
superimposed on an R-band CTIO 1.5~m telescope CCD image; the contour
threshold is determined as for Figure~\ref{hriconts}.  Two X-ray
sources in addition to J2310-437 are seen; the one to the SW may be
associated with a galaxy, whereas that to the SE may be associated
with a faint object apparent in the I-band image of
Tananbaum et al.~(1997).  Neither of these additional
X-ray sources was detected in the radio in our ATCA observations.
\label{hri2opt}}
\end{figure*}

\begin{figure*}
\plotone{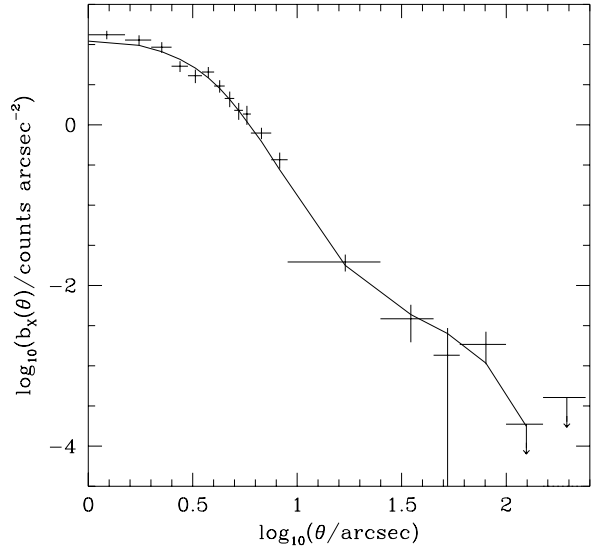}
\caption{The observed HRI radial profile from November 1995
(net counts and $1\sigma$ uncertainties plotted as crosses) gives an
excellent fit to a point source characterized by the nominal PRF of
David et al.~(1996); $\chi^2 = 13.7$ for 18 degrees
of freedom. Background, as measured in a source-centered annulus of
radii 2.5 to 4 arcmin, has been subtracted from the data, and the PRF
is normalized to the net counts in the profile.
\label{hri1profile}}
\end{figure*}

Aside from the effects of the aspect-correction residual errors in the
1996 HRI data, there is no obvious extended X-ray emission associated
with J2310-437; instead a bright central component is seen, and two
fainter sources are detected within a radius of 3 arcmin
(Fig.~\ref{hri2opt}).  The 1995 data give an excellent fit to the
nominal PRF of \markcite{david96}David et al.~(1996), as seen in
Figure~\ref{hri1profile}.

The HRI data constrain the models which \markcite{tan97}Tananbaum et
al.~(1997) fit to the ROSAT PSPC data.  The PSPC data give an {\it
acceptable\/} fit to a single component $\beta$ model, appropriate for
gas in hydrostatic equilibrium\footnote{The $\beta$-model notation
adopted here (e.g., \markcite{sar86}Sarazin~1986, and see
Table~\ref{clusterpars}) relates to the King-model formalism in
\markcite{tan97}Tananbaum et al.~(1997) via $p = 3\beta - 0.5$.}.
However, applying the corresponding best-fit model parameters to the
1995 HRI data gives $\chi^2_{\rm min} = 169$ for 17 degrees of
freedom, as compared with $\chi^2_{\rm min}$ = 14 for the fit to a
point source (18 degrees of freedom).

\begin{figure*}
\plotone{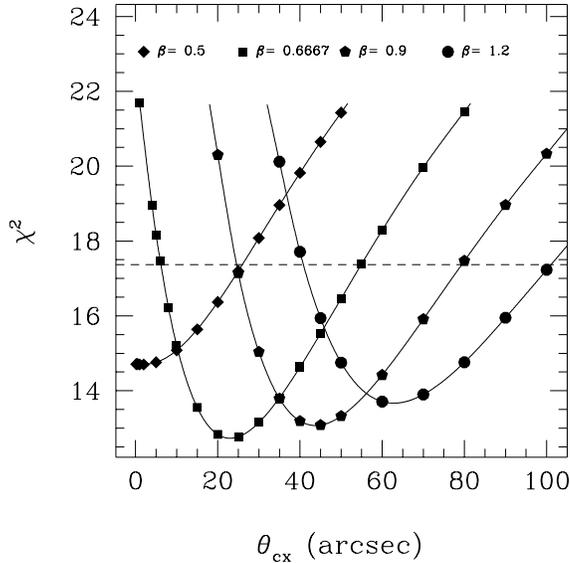}
\caption{$\chi^2$ vs core radius, $\theta_{\rm cx}$ for a fit
of the ROSAT PSPC radial profile (1172 net counts) in 14 bins out to a
radius of $210$~arcsec with a $\beta$-model plus point source.  The
$\beta$-model parameters for the best fit are $\beta = 2/3$ and
$\theta_{\rm cx} = 25$ arcsec.  The dashed line is at $\chi^2_{\rm
min} + 4.6$, corresponding to 90\% confidence for two interesting
parameters.  As an extension to the work of Tananbaum et al.~(1997),
the fraction of counts in the point source (within the extraction
radius), $f$, has been fitted rather than fixed at 0.8.  The maximum
permitted fraction ($f_{\rm max}$) for any $\beta$ is at the largest
permitted $\theta_{\rm cx}$ for that $\beta$.  We find that $f_{\rm
max}$ depends negligibly on the value of $\beta$ and is $\sim 0.9$ at
90\% confidence.  $f_{\rm min}$ for a given $\beta$ corresponds to the
smallest permitted $\theta_{\rm cx}$ for that $\beta$.  For example,
$f_{\rm min} = 0.5$ for $\beta = 2/3$, $\theta_{\rm cx} = 6$~arcsec,
and $f_{\rm min} = 0.76$ for $\beta = 1.2$, $\theta_{\rm cx} =
40$~arcsec.
\label{pspcchi}}
\end{figure*}

The PSPC data give an {\it unacceptable\/} fit to unresolved emission
alone, and so why is there no resolved
emission evident in the HRI data?  We have extended the PSPC model
fitting beyond that described in \markcite{tan97}Tananbaum et
al.~(1997) in order to examine two-component
fits ($\beta$-model plus unresolved emission) in which the relative
normalizations of the two components are allowed to vary.  Results are
shown in Figure~\ref{pspcchi}.  The best-fit model has $\beta = 2/3$
and core radius $\theta_{\rm cx} = 25$ arcsec, with roughly 80\% of
the counts in the unresolved component.  However, larger $\beta$ and
$\theta_{\rm cx}$ are allowed within 90\% confidence limits, and then
the emission in the unresolved component becomes closer to 90\%.

We have fitted the HRI radial profile to point-like plus resolved
emission, where the combination of parameters describing the resolved
emission (including normalization) is constrained by the PSPC fits
(Fig.~\ref{pspcchi}).  We allow the normalization of the point-like
component to be a free parameter, since this emission might plausibly
have varied in intensity between the epoch of the PSPC observations
(Nov.~1993) and that of the HRI.  Our spectral conversions use the
Galactic column density with a power law of energy index $\alpha =
1.43$ ($f_\nu \propto \nu^{-\alpha}$) for the point-like emission and
a Raymond-Smith spectrum of $kT = 1.18$~keV (50\% cosmic abundances)
for the $\beta$-model components (\markcite{tan97}Tananbaum et
al.~1997).  Although the two-component model giving the best fit to
the PSPC data is a relatively poor fit to the HRI data, there are models
within the 90\% confidence range for the PSPC which are acceptable
(Table~\ref{hripspc}).  The constraint on extended emission from the
HRI data places a lower limit of $\sim 0.36~\mu$Jy on the 1~keV flux
density of the unresolved emission measured with the PSPC, with a
preferred value of 0.4~$\mu$Jy.
We adopt $\beta = 2/3, ~\theta_{\rm cx} = 55$~arcsec as
the new best-fit $\beta$-model parameters.  

\begin{deluxetable}{cclccc}
\tablefontsize{\footnotesize}
\tablewidth{0pt}
\tablecaption{HRI radial fits to a PSPC-normalized $\beta$ model
plus a point source.
\label{hripspc}}
\tablehead{
\colhead{$\beta$}     & \colhead{$\theta_{\rm cx}$ (arcsec)}  &
\colhead{$t_{\rm cool}$ (yr) \tablenotemark{a}}&
\colhead{$f_{1~{\rm keV,~AGN}}$ ($\mu$Jy) \tablenotemark{b}}&
\colhead{$\left({\rm HRI \over PSPC}\right)_{\rm AGN}$  \tablenotemark{c}} & 
\colhead{$\chi^2$ for 17 dof} 
}
\startdata
0.5 & 2 &$3.2 \times 10^8$& $0.33 \pm 0.014$ & 
$1.19 \pm 0.08\phantom{5}$ & 25.5 \nl
0.5 & 25 &$1.3 \times 10^{10}$& $0.39 \pm 0.014$ & 
$1.09 \pm 0.065$ & 17.0 \nl
2/3 & 6 &$5.8 \times 10^8$& $0.22 \pm 0.014$ &
$1.51\pm 0.13\phantom{5}$ & 88.0 \nl
2/3 & 25 &$7.4 \times 10^9$& $0.37 \pm 0.014$ &
$1.13 \pm 0.07\phantom{5}$ & 26.6  \tablenotemark{d}\nl
2/3 & 55 &$2.7 \times 10^{10}$& $0.40 \pm 0.014$ &
$1.09 \pm 0.065$& 16.7  \tablenotemark{e}\nl
0.9 & 25 &$4.6 \times 10^9$& $0.33 \pm 0.014$ &
$1.21 \pm 0.08\phantom{5}$& 50.9 \nl
0.9 & 80 &$3.6 \times 10^{10}$& $0.40 \pm 0.014$ &
$1.08 \pm 0.065$& 16.9 \nl
1.2 & 100 &$4.3 \times 10^{10}$ & $0.40 \pm 0.014$ &$1.08 \pm 0.065$& 17.4 \nl
\tablenotetext{a}{Average cooling time within $\theta_{\rm cx}$. These
values should be compared with a Hubble time of $\sim 2 \times
10^{10}$ yr.}
\tablenotetext{b}{1 keV flux density for the unresolved component from
the PSPC fit, corrected for the fact that only 96\% of the PSPC counts
lie within a source-centered circle of radius 3.5 arcmin.}
\tablenotetext{c}{Ratio of normalization of the unresolved emission
in the HRI and PSPC fits, corrected for the response of the
instruments assuming a power-law
spectrum with $\alpha = 1.43$ and Galactic absorption.}
\tablenotetext{d}{Best-fit case for the PSPC data.}
\tablenotetext{e}{Adopted as giving new best-fit parameter values.}
\tablecomments{Figure~\ref{pspcchi} shows goodness of fits
to the PSPC data.}
\enddata
\end{deluxetable}

\subsection{X-ray Light Curve}

Table~\ref{hripspc} gives the ratio of the 1~keV flux densities for
the point component in the HRI and PSPC fits.  Although no value is
inconsistent with unity at high confidence (save the $\beta = 2/3$,
$\theta_{\rm cx} = 6$ arcsec model, which can be ruled out on the
basis of $\chi^2$), and values are closer to unity in cases where the
fits are more acceptable, there is a suggestion that the flux density
increased by about 10\% between the PSPC (Nov.~1993) and HRI
(Nov.~1995) observations.  Any hardening of the spectrum with time
would increase the ratio; for example, hardening to $\alpha = 0.65$
(closer to the extrapolated radio to X-ray spectral index, see later),
means values in column~5 of Table~\ref{hripspc} should be multiplied
by a factor of 1.4.  The HRI observations indicate no significant
change in intensity between Nov.~1995 ($0.078 \pm 0.0034$ cts
s$^{-1}$) and May 1996 ($0.071 \pm 0.002$ cts s$^{-1}$) and no
significant intra-month variability. However, the complexity of the
source and our rather imprecise knowledge of the extended structure
and source spectrum conservatively suggest that variability of the
point component by as much as 50\% over a timescale of a few years
cannot be ruled out.

\section{ATCA Radio Observations}

We observed J2310-437 with the CSIRO/ATNF Australia Telescope Compact
Array (ATCA) for 13 hours (including setup and calibration) between UT
1997 January 29th 22:30 and January 30th 11:30.  The array was in the
6A configuration, giving baselines from 0.34 to 6~km.  All six
antennas were on line for most of the time.  The observations were
made in four frequency bands (Table~\ref{radcores}), with
half the time at L and S bands, and half at C and X. Observing
intervals were interleaved to give full hour-angle coverage and no
large gaps in the uv-plane.  Heavy rain and thick cloud necessitated
phase and gain calibrations every 12 minutes.  Secondary calibrations
used the nearby quasar PKS~2311-452, while the primary flux-density
calibration is based on observations of PKS~1934-638 at the beginning
and middle of the 13-hour run.  32 frequency channels were used for
each band; 16 of these are independent, but one (channel 4) was
removed because of self interference, and the two end channels were
removed because of low signal strength.
10-second integrations were used for all observations at all bands.
The data were calibrated, mapped, and self-calibrated using the ATNF's
Miriad software package.  Final mapping and polarization analysis used
AIPS, with weighting schemes chosen to maximize signal-to-noise or to
permit the best spectral-index measurements.

\begin{deluxetable}{ccc}
\tablefontsize{\footnotesize}
\tablewidth{0pt}
\tablecaption{Radio-core flux densities at the four observing frequencies. 
\label{radcores}}
\tablehead{
\colhead{Band}     & \colhead{Frequency (MHz)}  &
\colhead{$f_{\rm core}$ (mJy)}
}
\startdata
L& 1384 & 27.7 $\pm$ 2.0 \nl
S& 2368 & 27.9 $\pm$ 0.6 \nl
C& 4800 & 20.6 $\pm$ 0.5 \nl
X& 8640 & 20.9 $\pm$ 0.3 \nl
\tablecomments{All bands were 112~MHz wide, centered on the reported
frequency, with an 8~MHz ``notch'' for a missing channel}
\enddata
\end{deluxetable}

\subsection{Radio Images and Polarization}

\begin{figure*}
\epsscale{0.8}
\plotfour{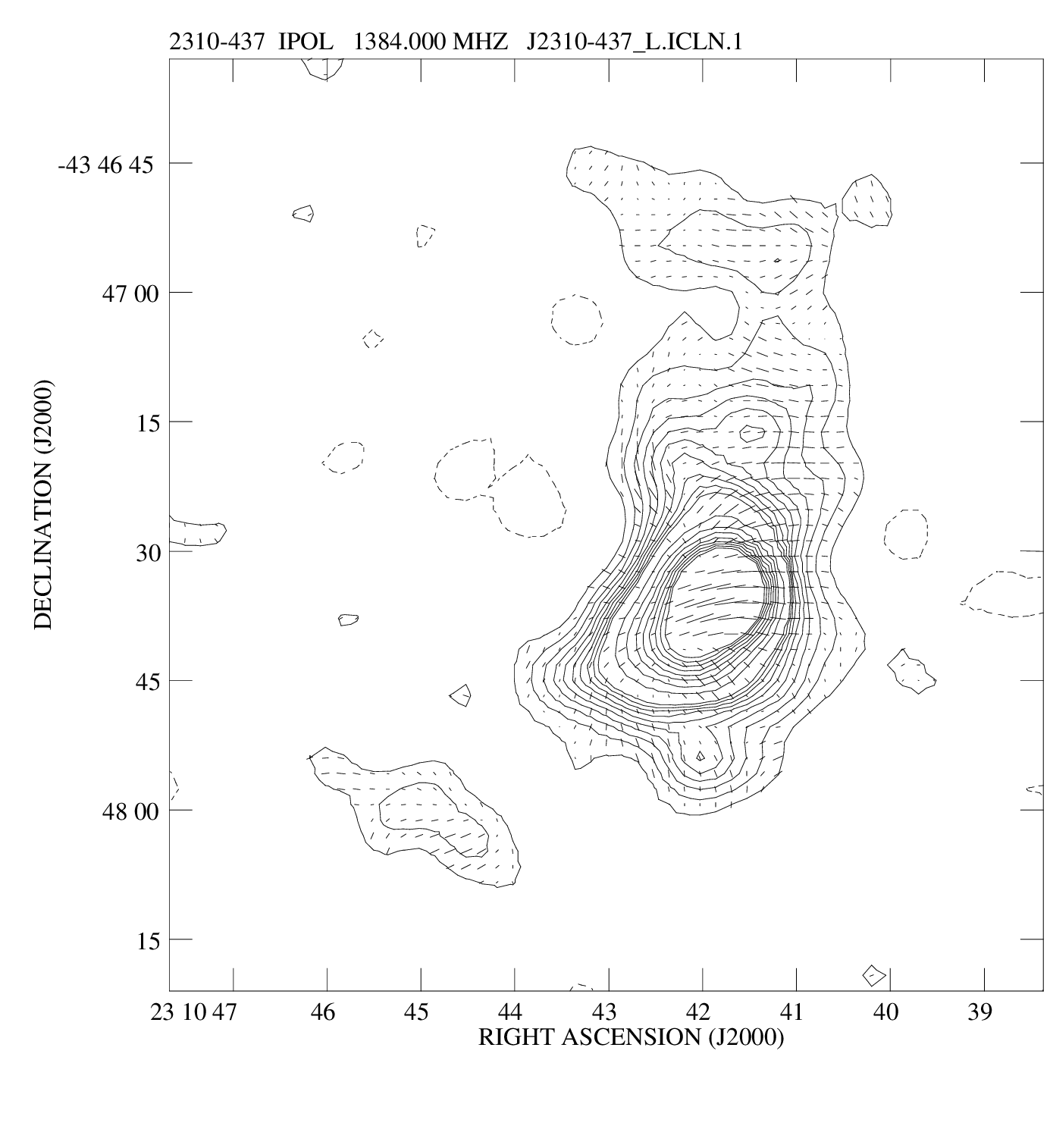}{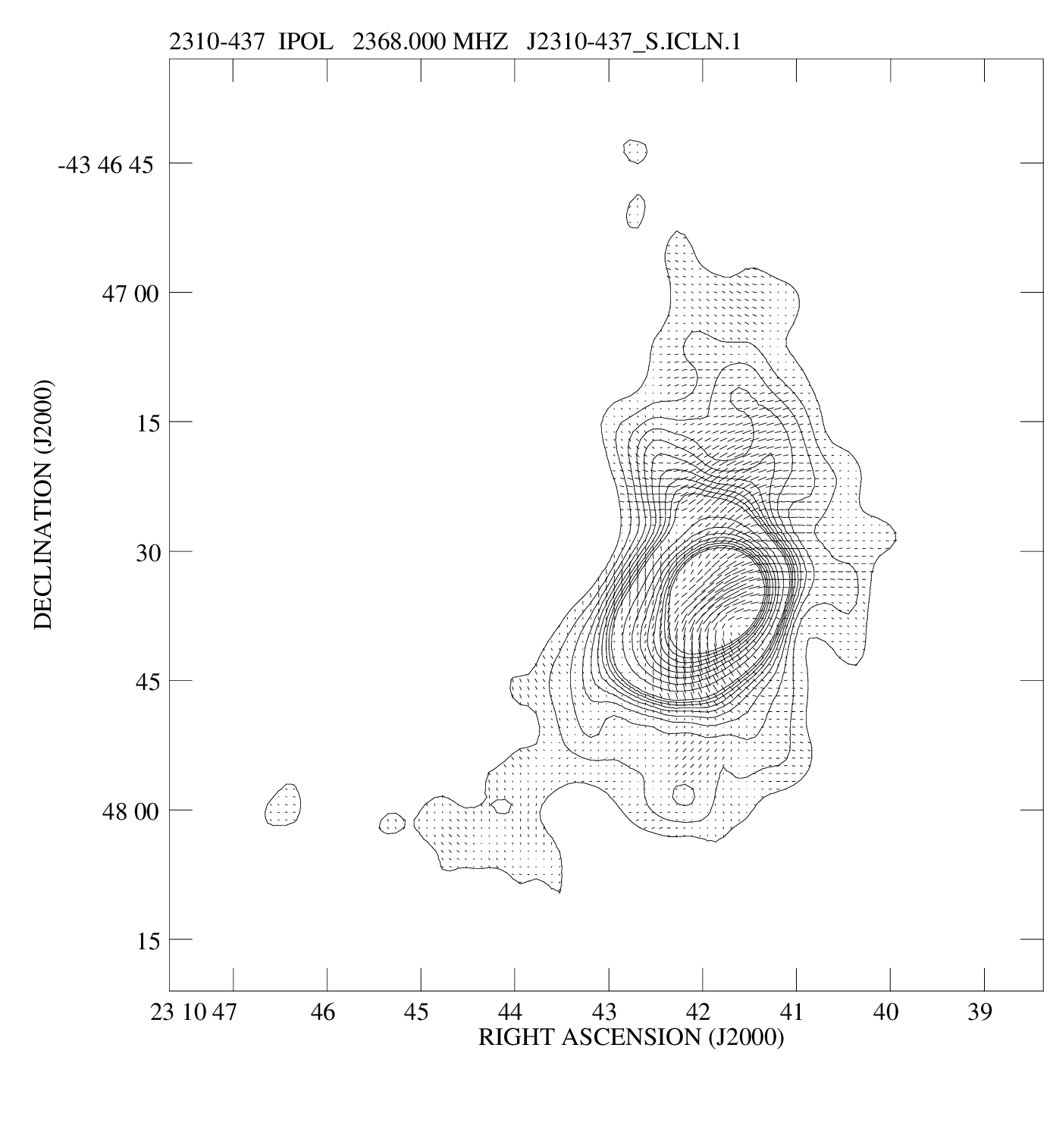}{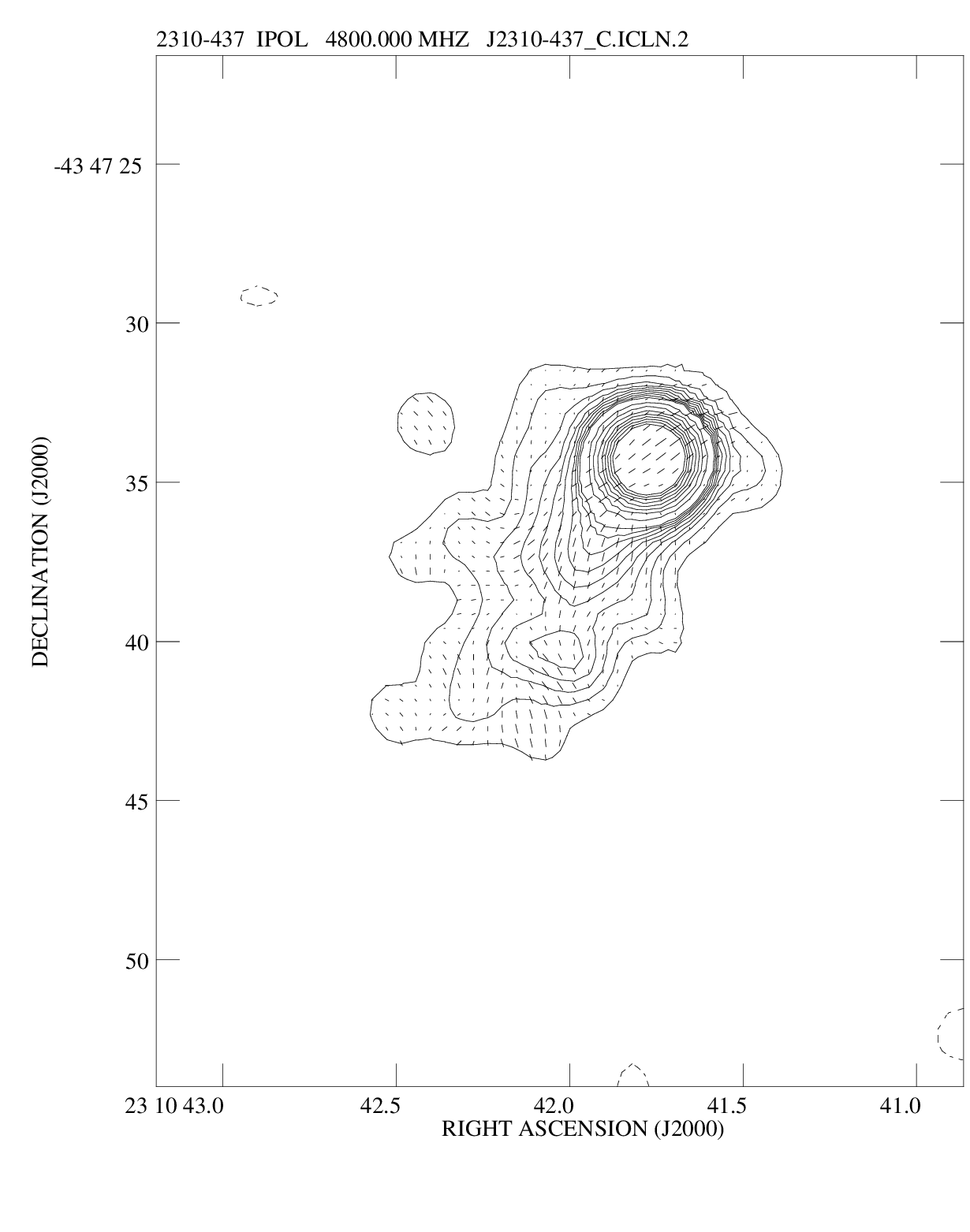}{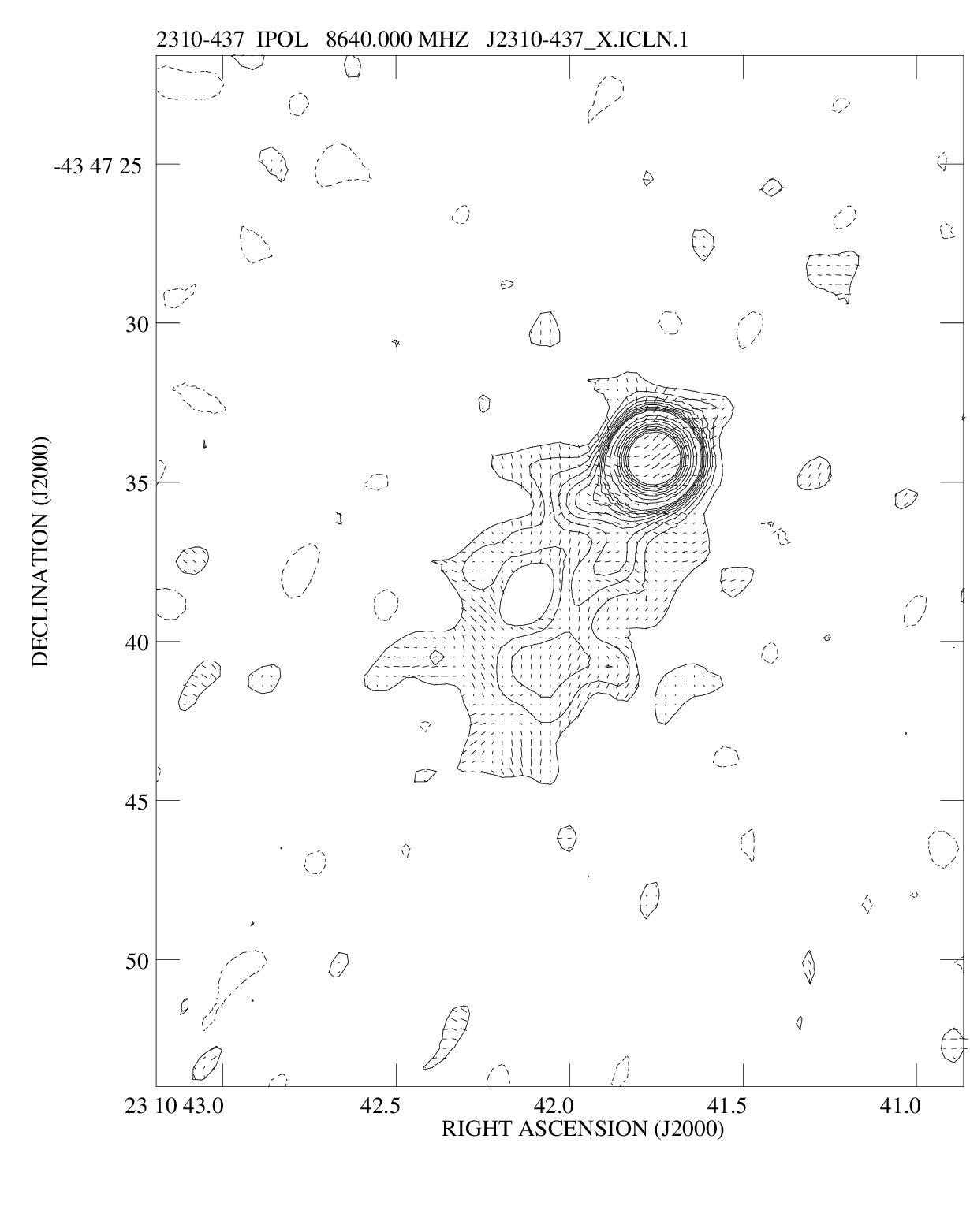}
\caption{L-band (top left), S-band (top right), C-band (bottom left)
and X-band (bottom right) radio maps of J2310-437 with polarization E
vectors (length proportional to polarized flux) superimposed. Note
scale change between upper and lower figures. The restoring Gaussian
beams have a FWHM of $6.5''$ (L-band and S-band), $2''$ (C-band) and
$1.5''$ (X-band).  Contours are at 0.2, 0.4, 0.6, 0.8, 1, 1.2, 1.4,
1.6, 1.8, 2, 3, 4, 5, 6, 7, 8, 9, 10 mJy, with -0.2 mJy shown dashed.
The maps show a transverse magnetic field following the line of the
jet and significant radio polarization in the core.  The signal to
noise of the polarization is low within the lowest three contours of
each map.
\label{radpol}}
\end{figure*}

Figure~\ref{radpol} shows contour maps of the processed data for each
frequency band with polarization E vectors superimposed.  The maps are
affected by uv-sampling and frequency; the L-band image provides our
best information about large-scale structure and the X-band measures
the core and inner jet-structure only.

\subsubsection{X Band}

The image shows one-sided extended emission to the SE with $16
\pm 3$ mJy of diffuse emission extending about 10 arcsec at a position
angle of $150^\circ$.  This jet appears to come out from the core at a
position angle of about $130^\circ$. Core polarization of about 2\% is
seen, with the magnetic field aligned perpendicular to the jet.
Polarization in the jet is detected at low signal to noise and appears
to peak at $\sim 15$\% about 2~arcsec SE of the nucleus, still with
the inferred magnetic field perpendicular to the jet.

\subsubsection{C Band}

The total flux in our measurements is $42 \pm 4$ mJy.  Since the PMN
survey measured 61~mJy (\markcite{griff93}Griffith \& Wright 1993),
about 20~mJy of C-band diffuse emission should be present on scale
sizes greater than 20 arcsec, which is the largest scale to which the
6A array is sensitive.  It is also possible that a variable core
contributes to the discrepancy.  The diffuse structure and jet are one
sided and resemble the X-band in structure and polarization.  There is
no significant Faraday rotation present between X and C bands.

\subsubsection{S Band}

The image shows two-sided structure with some evidence of tails
curving back to the N and ESE.  The polarization structure in the core
is particularly striking.  A small rotation of the plane of
polarization, about 15 degrees, is seen relative to the E vectors at C
band.

\subsubsection{L Band}

The curvature of the two-sided tail structure is most evident at this
frequency. The total measured flux density is about 99~mJy, and the
extent of the L-band image is roughly that of the cluster-related PSPC
X-rays.  The spectral index between the S and L bands in the SE and NE
lobes is $\alpha \sim 0.8$ ($f_\nu \propto \nu^{-\alpha}$), in the
normal range for optically-thin synchrotron emission.  The position
angle of the jet to the SE is similar to that in the inner region at
X-band, suggesting bending from a position angle of $\sim 130^\circ$
to $\sim 150^\circ$ and back to $\sim 130^\circ$ with increasing
distance from the core.  The E vectors in the core are rotated by
about 50 degrees with respect to those at S band, and the polarized
fraction peaks at about 30\%, to the NW of the core.  The Faraday
rotation measure, which is about $-25$~radians m$^{-2}$ (varying by
$\sim 10$ radians m$^{-2}$ over the inner part of the source), could
arise entirely from gas and fields in our Galaxy.

\subsection{The Radio Core}

The radio core lies at $\alpha = 23^{\rm h}$ $10^{\rm m}$ $41^{\rm
s}.76 \pm 0.01$, $\delta = -43^\circ$ $47'$ $34''.3 \pm 0.3$ (J2000).
The flux densities (Table~\ref{radcores}) imply a flat spectrum, with
a component of the emission becoming optically thin between the C and
S bands.  While the position confirms the association between the
radio emission, the optical galaxy, and the X-ray source, there are
remaining small discrepancies.  The galaxy centroid determined from
the Southern Sky Survey, as digitized by the Space Telescope Science
Institute, agrees in right ascension but is 1.6 arcsec north in
declination.  The most likely explanation is an error in the
calibration of the optical reference frame for this part of the sky.
The centroid of the HRI position agrees to within an arcsecond with
that from the digitized sky survey; on the one hand this is surprising
because the HRI aspect-correction uncertainties can result in
centroids which are in error by several arcseconds, but on the other
hand the closer agreement to the optical centroid than to the radio
position may reflect the fact that X-ray positions are determined with
respect to the optical reference frame rather than the radio. The
X-ray centroid for the lower-resolution PSPC data
(\markcite{tan97}Tananbaum et al.~1997) is 3.5 arcsec to the south of
the radio core.

\subsection{The Overall Radio Structure} \label{overallradio}

At X- and C-band the radio structure is predominantly one-sided.
Since at lower frequencies the structure is clearly two-sided, a
possible interpretation is that the source is a (slightly wiggly)
twin jet, where the SE side experiences greater Doppler boosting
towards the observer. The data allow us to place limits on the jet to
counter-jet ratio, $R$, of $> 3$ from the X-band map (with $R \sim
3.5$ on $2''$ scale), and $> 8$ on $5''$ scales from the C-band map.
At S- and L-band the resolution is too low to provide useful
information.

It is hard to classify this source by comparison with other radio
sources seen in poor clusters.  However, in section~\ref{agnradx} we
note that at least one radio source, the X-ray selected BL~Lac object
0548-322, has a similar morphology.  If the bending in J2310-437 which
is measured at L-band continues at lower surface brightness, the
source power and structure are somewhat reminiscent of narrow-angle
tail (NAT) sources such as 3C~83.1 B, in which the jets are bent back
to become approximately parallel.  However, such sources are usually
associated with non-dominant cluster galaxies, with the bending
attributed to motion of the host galaxy through the
intra-cluster medium.  Wide-angle tail (WAT) sources are more often
seen in the centers of clusters.  The prototypical WAT is 3C~465 in
Abell 2634, but WATs also occur in poor clusters, such as the source
1919+479 (\markcite{pink94}Pinkey, Burns \& Hill 1994).  The bending
of WATs is often interpreted as due to galaxy mergers.  WAT sources
are typically larger than 500 kpc in extent, whereas the L-band
emission we measure in J2310-437 spreads only over about 300 kpc.
Also the jets in WATs tend to show hot-spots at the position of
bending, rather than the smooth fall-off in intensity seen in
J2310-437.  J2310-437 is more reminiscent of low power radio galaxies
such as 3C~449 or 3C~31, but unlike them exhibits a
severely bent outer structure.  More sensitive X-ray measurements of
the cluster environment are needed before we can say if this bending
can be explained as a buoyancy effect or interaction with the hot
X-ray emitting gas.

\section{CTIO Optical Spectrum} \label{opticalspectrum} 

\begin{figure*}
\epsscale{1.0}
\plotone{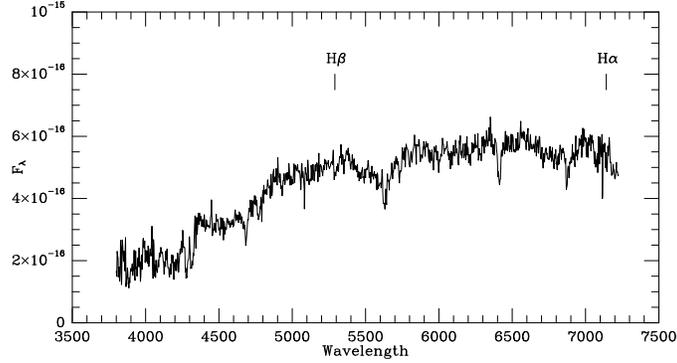}
\caption{Flux-calibrated observed-frame spectrum of the core of
J2310-437, obtained with the CTIO 1.5~m telescope.  Flux density, in
units of ergs cm$^{-2}$ s$^{-1}$ \AA$^{-1}$, is plotted against
wavelength in \AA.  There is no evidence for an AGN in the spectrum
from emission lines or the size of the Ca~II break (at 4350~\AA\
observed frame, 4000~\AA\ rest frame). The flux density is lower than
in the earlier spectrum of Tucker et al.~(1995) due to a different
slit size and extraction region.
\label{optspec}}
\end{figure*}

Our earlier optical spectrum of J2310-437 (\markcite{tuck95}Tucker et
al.~1995) did not cover H$\alpha$ or the Ca~II break, but both
regions can provide evidence for nuclear
activity.  Consequently, new observations were performed with the CTIO
1.5 meter telescope on 1996 May 15.  We used the Cassegrain
spectrograph with the Loral 1K CCD.  The wavelength range of 3800 --
7230~\AA\ was sampled with 8~\AA\ resolution (FWHM), using a slit
width of 3 arcsec. The exposure time was 1500~s during clear sky
conditions, and the subsequent target was a flux calibration standard
that was observed at a similar value of the air mass (1.20).
Spectral reductions and calibrations were performed using IRAF.  We
deliberately limited the extraction aperture to a window of $\pm 1.5$
arcsec, thereby sampling only the $3'' \times 3''$ core of the galaxy.
The result is shown in Figure~\ref{optspec}.  There is no H$\alpha$
emission line, at an upper limit of $6.0 \times 10^{-16}$ ergs
cm$^{-2}$ s$^{-1}$.  For comparison purposes, we also extracted a
spectrum from a source-centered $12'' \times 3''$ region.

The blue end of the $3'' \times 3''$ spectrum (Fig.~\ref{optspec})
shows normal absorption features, including the Ca~II H and K lines
and the g band, redshifted by z = 0.0886.  Following
\markcite{dress87}Dressler \& Shectman (1987) we quantify the size of
the Ca~II break as the fractional drop in flux ($f_\nu$ units) between
the 4050-4250~\AA\ and 3750-3950~\AA\ galaxy rest-frame bands. The
values found ($0.46 \pm 0.02$ for the $3'' \times 3''$ spectrum and
$0.47 \pm 0.02$ for the $12'' \times 3''$ spectrum) are both in the
normal range for elliptical galaxies.

Since there is no evidence of H$\alpha$ emission, we derive an upper
limit for the blue emission from any AGN nucleus by calculating the
maximum continuum that can be added to the $3'' \times 3''$ spectrum
before the Ca~II break is diluted to 0.40, the value proposed by
\markcite{mar96}March\~a et al.~(1996) as borderline between galaxies
with and without an AGN continuum, and a value $\sim 3\sigma$ away
from our results.  A convenient model for the continuum source is
provided by H1722+119 (\markcite{briss90}Brissenden et al.~1990),
which is an X-ray selected BL~Lac object (XBL) with no evidence for a
visible host galaxy. The optical continuum of H1722+119 is well fit by
a single power law with $f_\lambda \propto \lambda^{-0.12}$,
corresponding to $f_\nu \propto \nu^{-1.88}$.  In using this
relatively steep (i.e., red) spectrum rather than the average value of
$f_\nu \propto \nu^{-0.65}$ found (with large dispersion) for multiple
observations of 10 XBLs by \markcite{fal94}Falomo, Scarpa \&
Bersanelli (1994), we are being conservative by tending to maximize
the allowed nonthermal contribution to the spectrum of J2310-437 at
the Ca~II break.  After converting to the observed frame, for
frequencies close to the Ca~II break we find a maximum BL~Lac flux
density of $f_\lambda = 1.4 \times 10^{-16} (\lambda/{\rm
\AA})^{-0.12}$ ergs cm$^{-2}$ s$^{-1}$~\AA$^{-1}$; $f_\nu = 2.5 \times
10^{29} (\nu/{\rm Hz})^{-1.88} \mu$Jy.

Evidence that an AGN optical continuum may lurk at a level close to
our limit comes from comparing our results with
\markcite{cacc97}Caccianiga \& Maccacaro (1997)'s recent spectrum
taken using the ESO 3.6~m telescope with a 1.5 arcsec slit.  From an
extraction region of $0.6'' \times 1.5''$ (Maccacaro 1997; private
communication), containing less galaxy light than our $3'' \times 3''$
spectrum, they find a Ca~II break of $0.38\pm 0.04$ compared with our
value of $0.46\pm 0.02$.  Although the values are consistent to within
$2\sigma$, they suggest that a central continuum may dilute the break.
When the spectral extraction region excludes the inner $1.8'' \times
1.5''$ of the galaxy, \markcite{cacc97}Caccianiga \& Maccacaro (1997)
find a contrast of $0.47\pm 0.05$, in good agreement with our results.
A detailed comparison of the allowed level of nonthermal continuum is
not possible, since the \markcite{cacc97}Caccianiga \& Maccacaro
spectrum has no absolute flux calibration, and different slit widths
and extraction apertures are used.  However, their measurement is
consistent with roughly equal AGN and elliptical galaxy contributions
at $\sim$~4300~\AA\ (observed frame) if the elliptical galaxy has a
normal Ca~II break of 0.6.  This result may be in agreement with our
upper limit of about 20\% AGN continuum in the same wavelength region
due to our larger slit size and extraction region.  Further
improvement in the optical evidence for an AGN would require a
high-quality UV spectrum taken with a narrow slit or a high-quality
image, such as is possible with the HST.

\section{Discussion}

\subsection{The Cluster}\label{clustersection}

The X-ray source includes a component from a hot atmosphere of cluster
scale with $\beta = 2/3$ and $\theta_{\rm cx}$ = 55~arcsec
(section~\ref{xrayspatial}).  Physical parameters for the cluster gas
are given in Table~\ref{clusterpars}. The 0.5 - 2.5~keV
luminosity is in good agreement with Abell richness class 0 clusters
(\markcite{briel93}Briel \& Henry 1993), and the extrapolated 2 -
10~keV luminosity of $3.7 \times 10^{42}$ ergs s$^{-1}$ is consistent
with \markcite{david93}David et al.~(1993)'s temperature - luminosity
correlation.

\begin{deluxetable}{lll}
\tablefontsize{\footnotesize}
\tablecaption{Parameters for the Cluster X-ray Emission around J2310-437. 
\label{clusterpars}}
\tablewidth{0pt}
\tablehead{
\colhead{Parameter}     & \colhead{Value} & \colhead{Reference} 
}
\startdata
Redshift & 0.0886 & 1\nl
Galactic $N_H$ & $1.5 \times 10^{20}$ cm$^{-2}$ & 2 \nl
$\beta$ \tablenotemark{a}& 2/3  & 3\nl
$\theta_{\rm cx}$ \tablenotemark{a}& 55~arcsec & 3\nl
$B_o$ \tablenotemark{a} & $(2.84 \pm 0.49) \times 10^{-6}$ PSPC cts
s$^{-1}$ arcsec$^{-2}$ (0.17-1.8 keV)
& 3\nl
Total count rate & $0.054 \pm 0.01$ PSPC cts s$^{-1}$ (0.17-1.8 keV) & 3 \nl
$kT$, Abundance & 1.18~keV, 50\% cosmic & 4 \nl
Central density & $1.9 \times 10^{-3}$ cm$^{-3}$ & 3 \nl
$L_{0.1 - 2.4~\rm keV}$ & $2.6 \times 10^{43}$ ergs s$^{-1}$ & 3 \nl
$L_{0.5 - 2.4~\rm keV}$ & $1.8 \times 10^{43}$ ergs s$^{-1}$ & 3 \nl
\tablenotetext{a}{ For $\beta$ model of the form $S = B_o
(1 + {\theta^2 \over \theta_{\rm cx}^2})^{0.5 - 3 \beta}$}
\tablerefs{(1) \markcite{tuck95}Tucker et al.~1995; 
(2) \markcite{tan97}Tananbaum et al.~1997 based on
\markcite{heil79}Heiles \& Cleary 1979; (3) this paper; (4) ROSAT
PSPC from \markcite{tan97}Tananbaum et al.~1997}
\enddata
\end{deluxetable}

Table~\ref{hripspc} gives the mean cooling time for gas within
$\theta_{\rm cx}$ calculated using equations 10 and 25 of
\markcite{birkw93}Birkinshaw \& Worrall (1993). The cooling time
within one core radius is comparable to the Hubble time for the
best-fit model, suggesting that a strong cluster cooling flow is
unlikely. However, there are acceptable fits with small core radii and
significant cooling flows.  Whereas a $\beta$-model profile is quite
flat at small radii, a cooling flow would give excess emission of
steeper gradient.  A strong contribution of such a component of
scale-size $\gtrsim 5$~arcsec is ruled out by the excellent fit of the
PRF to the November 1995 HRI data (Fig.~\ref{hri1profile}).  Moreover,
since ROSAT finds compact X-ray emission (section~\ref{agnsection}) of
almost ten times the luminosity of the cluster, it is highly unlikely
that a cluster cooling flow dominates the central X-ray
emission. However, data of higher spatial resolution are needed to
place useful constraints on a possible contribution.

Before concluding that the predominant X-ray emission from J2310-437
is associated with an AGN, we should also consider X-ray emission from
unusually dense, sub-galaxy, hot gas.  For simplicity we characterize
the unresolved HRI emission as produced in a sphere of uniform density
with a radius of 11.4 kpc, corresponding to a conservative limit of
5~arcsec radius.  The two-component thermal fits to the PSPC data
(\markcite{tan97}Tananbaum et al.~1997) constrain the temperature of
the cooler gas to $\sim 0.1$~keV, giving a cooling time of $\sim 1.7
\times 10^7$ years and $\sim 6500 M_\odot$ per year of cooling gas.
We might expect one H$\alpha$ recombination photon to be released per
cooling atom, giving a rate of $\sim 2 \times 10^{53}$ H$\alpha$
photons s$^{-1}$.  However, our upper limit of $6.0 \times 10^{-16}$
ergs cm$^{-2}$ s$^{-1}$ for the H$\alpha$ flux
(section~\ref{opticalspectrum}) corresponds to $\lesssim 10^{51}$
H$\alpha$ photons s$^{-1}$.  Even assuming that the unresolved HRI
emission is associated with the warmer PSPC component, at a
temperature of 1~keV, we still estimate a cooling rate of $\sim 600
M_\odot$ per year and many more H$\alpha$ photons than allowed by our
upper limit.  Moreover, in this case we would have the unusual
situation of the cooling component having a higher temperature than
the gas feeding it.  The problem is compounded by the fact that the
H$\alpha$ measured from cooling flows is typically 10 to even 1000
times larger than predicted by the simple one photon per recombination
argument, implying additional excitation processes
(\markcite{don97}Donahue \& Voit 1997). We therefore conclude that it
is extremely unlikely that the compact X-rays have a hot-gas origin.

Table~\ref{pressure} compares the minimum internal pressure, $P_{\rm
min}$, for features in the radio source with the pressure from the
X-ray emitting cluster gas.  We find that the gas is over-pressured
with respect to $P_{\rm min}$, as is common for kpc-scale structures
of low-power radio sources (e.g., \markcite{morg88}Morganti et
al.~1988, \markcite{bohr93}B\"ohringer et al.~1993,
\markcite{truss97}Trussoni et al.~1997).  Possible explanations (e.g.,
\markcite{hard98}Hardcastle, Worrall \& Birkinshaw 1998b) include
dynamical effects, where the outer radio structure is expanding, or
contributions to the jet pressure from relativistic or thermal
(entrained) protons, or clumping.  X-ray maps of higher resolution
than from the PSPC are required for a more detailed assessment of the
relationship between the rather unusual outer radio structure
(section~\ref{overallradio}) and the X-ray emitting cluster medium.

\begin{deluxetable}{cccccc}
\tablefontsize{\footnotesize}
\tablecaption{Minimum Pressure in L-band Radio Features and X-ray Gas Pressure.
\label{pressure}}
\tablewidth{0pt}
\tablehead{
\colhead{$\theta$ \tablenotemark{a}}  & 
\colhead{pa  \tablenotemark{a}} & 
\colhead{FWHM } &
\colhead{$B_{\rm eq}$ \tablenotemark{b}} &
\colhead{$P_{\rm min}$  \tablenotemark{b}} &
\colhead{$P_{\rm gas}$  \tablenotemark{c}}
\\
\colhead{(arcsec)}  & 
\colhead{ (degs) } & 
\colhead{(sq arcsec)} &
\colhead{($\mu$Gauss)} &
\colhead{ (dynes cm$^{-2}$) } &
\colhead{ (dynes cm$^{-2}$) }
}
\startdata
40 & 130 & $31 \times 14$ & 3.4 & $3.1 \times 10^{-13}$ & 
$5.2 \times 10^{-12}$ \nl
40 & 355 & $25 \times 14$ & 3.9 & $4.0 \times 10^{-13}$ & 
$5.2 \times 10^{-12}$ \nl
75 & 15 & $43 \times 37$ & 1.8 & $8.6 \times 10^{-14}$ & 
$2.5 \times 10^{-12}$ \nl
\tablenotetext{a}{Angular distance and position angle of feature from
the radio core}
\tablenotetext{b}{Equipartition magnetic field and minimum pressure in
radio feature
calculated assuming an electron spectral index of 2.6 extending from $3
\times 10^7$ to $2 \times 10^{11}$~eV, a plasma filling-factor of
unity, and no protons.
}
\tablenotetext{c}{Gas pressure using cluster parameters from Table~\ref{clusterpars}}
\enddata
\end{deluxetable}

\subsection{The AGN} \label{agnsection}

\subsubsection{Radio morphology and X-ray strength} \label{agnradx}

The most striking features of the radio maps of J2310-437 are

\begin{enumerate}

\item the flat-spectrum radio core, seen in the C- and
X-band maps; 
\item the one-sided, small-scale jet, seen best in the
C- and X-band maps; and
\item the two-sided large-scale plumes, seen best in the L-band,
low-resolution image.

\end{enumerate}

The immediate inference from (1) is that J2310-437, despite its
``normal'' optical appearance and line-free optical spectrum, does
indeed host an AGN.  Much of the interest in studying this object then
shifts to attempting to understand what type of AGN we are observing.

The large-scale, two-sided radio morphology of J2310-437 seen in the
low-resolution images is reminiscent of the structures of
\markcite{fan74}Fanaroff and Riley (1974) type 1 (FR1) radio galaxies
-- ``plumes'' of radio emission extending to either side of the bright
central part of the source near the radio core, and with no sign of
``hot-spots'' in the radio lobes. At high resolution the source
exhibits a 1-sided jet as often seen in FR1 radio galaxies. The
relatively low L-band radio power of the source, $P_{1384} \approx 3.5
\times 10^{31}$ ergs s$^{-1}$ Hz$^{-1}$, is also consistent with its
classification as an FR1 radio galaxy (\markcite{ow91}Owen \& White
1991).

However, the associated X-ray emission, at $l_{\rm 1~keV} = 1.5
\times 10^{26}$ ergs s$^{-1}$ Hz$^{-1}$, is anomalously bright for FR1
radio galaxies; a representative sample of such radio galaxies have
$l_{\rm 1~keV}$ in the range $6 \times 10^{22}$ to $10^{25}$ ergs
s$^{-1}$ Hz$^{-1}$ (\markcite{worr97}Worrall 1997).  The core X-ray
power of J2310-437, $P_{0.1 - 2.4 \rm keV} \approx 1.7 \times 10^{44}$
ergs s$^{-1}$, is more similar to that of a BL~Lac object.  But if we
interpret J2310-437 as a BL~Lac object embedded in an elliptical
galaxy, we must ask

\begin{enumerate}

\item Is the radio structure similar to that of other X-ray selected
BL Lacs (XBLs)?
\item  Why do we not see a classically-bright optical nucleus
of BL~Lac type?
\item Are there other objects of similar type, largely
unrecognized in existing samples of X-ray or radio sources?

\end{enumerate}

While radio-selected BL~Lacs are dominated by bright, variable, and
polarized radio nuclei, and have featureless optical spectra, XBLs
have lower radio powers and are generally found in lower-redshift
galaxies (\markcite{schw89}Schwartz et al.~1989). For a sample of 20
XBLs from the HEAO-1 A1 survey, \markcite{laur93}Laurent-Muehleisen et
al.~(1993) find an average total radio power $P_{\rm L-band} \approx
10^{31}$ ergs s$^{-1}$ Hz$^{-1}$, an average core/extended L-band
emission ratio $\approx 4$, and an average $P_{\rm 2 keV}/P_{\rm
L-band}$ ratio $\approx 10^{-4}$.  In J2310-437, the total radio power
($\approx 3.5 \times 10^{31}$ ergs s$^{-1}$ Hz$^{-1}$) lies close to
the expected value, the radio core is weak (the core/extended L-band
ratio is $< 0.25$), and the ratio of the X-ray and extended radio
powers is low at $\approx 2 \times 10^{-6}$, although not as low as
the average value of $\approx 3 \times 10^{-7}$ found by
\markcite{laur93}Laurent-Muehleisen et al.~(1993) for a sample of FR1
radio galaxies.  J2310-437 lies within the range of properties seen
for the XBLs, but with radio and X-ray cores which are at the weakest
end of the distribution with respect to extended radio power.  This
may suggest that the core of J2310-437 is less subject to relativistic
boosting than an average XBL.

0548-322 is a weak-cored XBL in the HEAO A1 sample; it has been mapped
by \markcite{ant84}Antonucci \& Ulvestad (1984) and
\markcite{laur93}Laurent-Muehleisen et al.~(1993). This source has
several similarities to J2310-437: $z =0.0689$, $P_{\rm L-band} = 8
\times 10^{31}$ ergs s$^{-1}$ Hz$^{-1}$, core/extended L-band ratio of
0.2, and $P_{\rm 2 keV}/P_{\rm L-band} \approx 4 \times 10^{-6}$.  The
radio structures and polarization are also similar: 0548-322 shows
large-scale radio plumes, extending over more than 150~kpc, and the
central structure suggests that a one-sided jet may be present. The
linear polarization is $\sim 10$~per cent in the core and $\sim 8$~per
cent in the extended emission, with a transverse magnetic field in
both regions. The optical field of 0548-322 indicates that, like
J2310-437, it is embedded in a cluster (in this case Abell richness
class 2; \markcite{fal95}Falomo, Pesce \& Treves 1995), and the
absolute magnitude of the host galaxy of 0548-322 ($M_{\rm V} =
-23.4$) is brighter than J2310-437 ($M_{\rm V} = -22.67$;
\markcite{tuck95}Tucker et al.~1995) by a factor similar to the ratio
of total radio powers.

The main difference between the two objects is the presence of strong
optical continuum in 0548-322, where $\sim 50$~per cent of the
4200~\AA\ light (observed frame) comes from the AGN
(\markcite{fal94}Falomo et al.~1994). This is large, particularly
considering the fact that the effective aperture used in the
observations was $8'' \times 8''$ and so dilution due to galaxy light
should be relatively higher than in our observation of J2310-437 for
which $\lesssim 28$~per cent of the 4200~\AA\ light can originate in
an AGN.  Through its radio and X-ray resemblance to 0548-322, we
interpret J2310-437 as a BL~Lac embedded in an elliptical galaxy.  We
now consider in more detail why we do not see an optical nucleus of
the BL~Lac type, and just how unusual J2310-437 might be.

\subsubsection{The weak optical nucleus} \label{agnopt}

\begin{figure*}
\epsscale{1.0}
\plotone{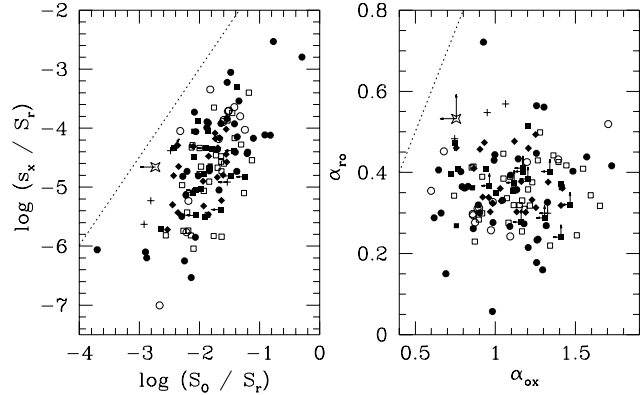}
\caption{Data for J2310-437 from Table~\ref{agnpars} (star) and
sources from Table~\ref{otherpars} (crosses) are
plotted with those for the 53 RGB XBLs (including those with low Ca~II
contrast) from Laurent-Muehleisen et al.~(1998; squares), the complete
EMSS XBL sample of 22 sources from Morris et al.~(1991; diamonds) and
the complete N-1ES sample of 48 XBLs from Perlman et al.~(1996a;
circles).  Open symbols show XBLs with no measured redshift.  The
left plot shows logarithmically the 1~keV X-ray ($S_{\rm x}$) vs
4400~\AA\ optical ($S_{\rm o}$) flux density, both normalized by the
5~GHz radio flux density ($S_{\rm r}$).  All flux densities are
corrected to the rest frame.  The median sample redshifts of 0.13
(RGB) and 0.15 (N~1ES) are used for sources without a redshift.
K-corrections for the XBLs use $\alpha_{\rm r} = 0$, $\alpha_{\rm o} =
1.0$, and $\alpha_{\rm x} = 1.2$ (RBL) or 1.35 (N-1ES).  X-ray fluxes
and K-corrections for the EMSS sample are computed from the ROSAT
spectra presented by Perlman et al.~(1996b).  Optical flux densities
are computed assuming the 4400~\AA\ flux density of a zero magnitude
object is $10^{6.65}$~mJy for O-band magnitudes for the RGB and N-1ES
samples, and $10^{6.463}$~mJy for V-band magnitudes for the EMSS
sample.  The right hand plot shows the same data in terms of
$\alpha_{\rm ro}$ vs $\alpha_{\rm ox}$, where these parameters are as
defined in Table~\ref{agnpars}.  The dashed line on both plots is the
locus of sources with a single spectral slope through the radio,
optical and X-ray.  J2310-437 is an outlier compared with XBL
populations and could lie on this locus.
\label{aoxplot}}
\end{figure*}

Multifrequency parameter values for J2310-437 are summarized in
Table~\ref{agnpars}.  \markcite{tan97}Tananbaum et al.~(1997) compared
earlier limits for two-point spectral indices $\alpha_{\rm ro}$ and
$\alpha_{\rm ox}$ with AGN from the work of \markcite{stock90}Stocke
et al.~(1990), and concluded that although the best overall match was
with BL~Lac objects, J2310-437 was either an extreme member of the
population or distinct.  In \markcite{tan97}Tananbaum et al.~(1997)
the two-point spectral indices are referenced to 5 GHz, 2500~\AA\ and
2~keV in the rest frame, whereas in this paper we reference them to 5
GHz, 4400~\AA\ and 1~keV (rest frame) for better comparison with
a larger selection of recent data for BL~Lac objects and to reduce
extrapolation errors. Our improved upper limit on AGN optical light
shifts J2310-437 to a more extreme region in ($\alpha_{\rm ro}$,
$\alpha_{\rm ox}$) parameter space (Fig.~\ref{aoxplot}).  XBLs tend to
show some degree of downwards curvature in the two-point spectrum
between the radio to X-ray, but J2310-437 is consistent with a single
power law of energy index $\alpha_{\rm rx} = 0.61$.  The two point
radio to X-ray spectral index for the XBLs is very similar to that for
J2310-437; the 123 XBLs in Figure~\ref{aoxplot} have a mean
$\alpha_{\rm rx}$ of $0.60 \pm 0.01$ and a median of 0.6 (this
compares with steeper average values of $0.86 \pm 0.08$ for a complete
sample of radio-selected BL~Lac objects [Sambruna, Maraschi \&
Urry~1996] and $\approx 0.85$ for the cores of low-power radio
galaxies [Worrall 1997], possibly because of the way the objects are
selected).  So, J2310-437 is not unusual compared with XBLs in its
ratio of core X-ray to radio emission, but it is extreme in emitting
so little optical continuum that a broad-band spectral break is not
required.

\begin{deluxetable}{lll}
\tablefontsize{\footnotesize}
\tablecaption{Parameters for the AGN in J2310-437. 
\label{agnpars}}
\tablewidth{0pt}
\tablehead{
\colhead{Parameter}     & \colhead{Value} & \colhead{Reference} 
}
\startdata
Redshift & 0.0886 & 1\nl
Galactic $N_H$ & $1.5 \times 10^{20}$ cm$^{-2}$ & 2 \nl
$\alpha_r$ for radio core & $\sim 0$ & 3\nl
$l_{5~\rm GHz}$ for radio core &
$(7.0 \pm 0.1) \times 10^{30}$ ergs s$^{-1}$ Hz$^{-1}$ & 3\nl
$f_{4400~\rm \AA}$ & $< 32$  $\mu$Jy & 3 \nl
$l_{4400~\rm \AA}$ ($\nu = 6.82 \times 10^{14}$ Hz) & 
$\leq 1.3 \times 10^{28}$ ergs s$^{-1}$ Hz$^{-1}$ & 3\nl
X-ray count rate & $0.086 \pm 0.004$ HRI cts s$^{-1}$ & 3 \nl
$f_{1~\rm keV}$  & $0.4 \pm 0.04 ~\mu$Jy & 3 \nl
$\alpha_x$  & 1.43 & 4 \nl
$l_{1~\rm keV}$ ($\nu = 2.4 \times 10^{17}$ Hz) & $1.5 \times 10^{26}$
ergs s$^{-1}$ Hz$^{-1}$ & 3 \nl
$L_{0.1 - 2.4~\rm keV}$ & $1.7 \times 10^{44}$ ergs s$^{-1}$ & 3 \nl
$\alpha_{\rm ox}$\tablenotemark{a} & $\leq$ 0.76 & 3\nl
$\alpha_{\rm ro}$\tablenotemark{b} & $\geq$ 0.53 & 3\nl
$\alpha_{\rm rx}$\tablenotemark{c} & 0.61 & 3\nl
\tablenotetext{a}{$\alpha_{\rm ox} = \log\ (l_{4400~\rm
\AA} /l_{1~\rm keV}) \ /\ 2.5465$}
\tablenotetext{b}{$\alpha_{\rm ro} = \log\ (l_{5~\rm GHz} /l_{4400~\rm
\AA}) \ /\ 5.135$}
\tablenotetext{a}{$\alpha_{\rm rx} = \log\ (l_{5~\rm
GHz} /l_{1~\rm keV}) \ /\ 7.68$}
\tablerefs{(1) \markcite{tuck95}Tucker et al.~1995; 
(2) \markcite{tan97}Tananbaum et al.~1997 based on
\markcite{heil79}Heiles \& Cleary 1979; (3) this paper; (4) ROSAT
PSPC from \markcite{tan97}Tananbaum et al.~1997}
\tablecomments{Flux densities relate to the observed frame.
Luminosities are for the rest frame. The error in the X-ray
flux density incorporates uncertainties in the amount of cluster
emission but does not include spectral-shape uncertainties.}
\enddata
\end{deluxetable}

Data for four sources with multifrequency spectra that resemble
J2310-437, most particularly in having weak or undetected optical
continuum coupled with similar X-ray to radio ratios, are compiled in
Table~\ref{otherpars} and plotted in Figure~\ref{aoxplot}.  In
addition, eight of the sources from the RASS-Green Bank (RGB) XBL
sample have seemingly normal elliptical galaxy spectra and are treated
as candidate XBLs by \markcite{laur98}Laurent-Muehleisen et
al.~(1998); they are plotted as limits in Figure~\ref{aoxplot}.
Sources with limits share J2310-437's consistency with a
single-component power law from the radio to X-ray, but none suggests
it as convincingly as J2310-437 itself.  Moreover, for J2310-437 a
hot-gas origin for the compact X-ray component is extremely unlikely
(section~\ref{clustersection}) whereas it remains a possibility for
the eight RGB sources (\markcite{laur98}Laurent-Muehleisen et
al.~1998), and indeed for similar candidates reported earlier by
\markcite{brink94}Brinkmann, Siebert \& Boller~(1994) and
\markcite{moran96}Moran et al.~(1996).  It will become clear whether
or not objects like J2310-437 are truly common or rare through further
optical, radio and X-ray follow-up of AGN selected from radio and
X-ray surveys (e.g., \markcite{laur98}Laurent-Muehleisen et al.~1998,
\markcite{cacc97}Caccianiga \& Maccacaro 1997).

\begin{deluxetable}{llccrrrc}
\tablefontsize{\footnotesize}
\tablewidth{0pt} \tablecaption{Parameters for optically deficient
nuclei possibly similar to J2310-437.
\label{otherpars}}
\tablehead{
\colhead{Name}     & \colhead{\ \ $z$}  & 
\colhead{$f_{\rm 1~keV}~(\mu$Jy)} &  \colhead{$\log L_{\rm rx}$ (ergs s$^{-1}$)} &
\colhead{$\alpha_{\rm ro}$} &\colhead{$\alpha_{\rm ox}$} &
\colhead{$\alpha_{\rm rx}$} &
\colhead{references}
}
\startdata
3C 264 & 0.0215 & 0.45 & 43.37 & 0.57  & 1.06 & 0.73 & 1,2,3 \nl
E 0336-248 & 0.251 & 0.45 & 45.12 & $\leq 0.48$ & 0.75 & $\leq 0.57$ & 4 \nl
26W20 & 0.054 & 0.63 & 44.75 & $\geq 0.3\phantom{0}$ & $\leq 1.33$ & 0.64 & 5, 6 \nl
PKS 2316-423 & 0.0549 & 0.85 & 44.27 & 0.55 & 0.95 & 0.68 & 7, 8, 6 \nl
J 2310-437 & 0.0886 & 0.4 & 44.12 & $\geq 0.53$ & $\leq 0.76$ & 0.61 & 9 \nl
\tablerefs{(1) \markcite{giov90}Giovannini et al.~1990; 
(2) \markcite{crane93}Crane et al.~1993; (3) \markcite{tan97}Tananbaum
et al.~1997; (4) \markcite{halp97}Halpern et al.~1997;
(5) \markcite{harris84}Harris et al.~1984;
(6) \markcite{silv98}Silverman et al.~1998;
(7) \markcite{wright97}Wright et al.~1997;
(8) \markcite{craw94}Crawford \& Fabian~1994; (9) this paper
}
\tablecomments{$L_{\rm rx}$ is the luminosity from 5~GHz to 1~keV in
the source frame assuming $\alpha_{\rm ro}$ and $\alpha_{\rm ox}$.
For $\alpha_{\rm ox} < 1$\ the X-ray contribution dominates and quoted
luminosities extend only to 1~keV.
Parameters for 3C 264 differ from those in Tananbaum et
al.~(1997) because of the redefinition of reference frequencies
for the two-point spectral indices and use here of an improved (lower) radio
core flux density of 200~mJy.  The radio flux density for E
0336-248 is treated as an upper limit due to lack of high spatial
resolution mapping.   For 26W20 the X-ray flux density is from the
average of the ROSAT measurements, and the optical is dominated by
galaxy light; no continuum separation has been attempted in references
5 and 6.  For PKS 2316-423 the higher spatial resolution mapping in
reference 7 updates information in reference 8.  The claimed
similarity to J2310-437 is in the last reference for each source.
}
\enddata
\end{deluxetable}

\subsubsection{A photon-starved jet?} \label{agnjet}

Since a single power law connecting the radio and X-ray cores runs a
factor of about 2 below the optical upper limit, it is possible to
interpret all of this emission as arising from a single region of
synchrotron-emitting plasma, where the flat radio spectrum is
attributed to a series of self-absorbed components and the
optically-thin synchrotron spectral index is $\alpha_s = 0.61$.
Assuming Galactic $N_H$, the best-fit power-law spectral index to the
PSPC X-ray emission of 1.43 (\markcite{tan97}Tananbaum et al.~1997) is
steeper than $\alpha_s$; fixing the X-ray spectral index at 0.61 gives
$\chi^2 = 29$ for a fit to the PSPC spectral data as compared with a
minimum $\chi^2$ of 18 for 28 degrees of freedom.  This result may
suggest that an energy-loss break in the electron spectrum is showing
up in the X-ray emission.  However, the quality of the data and the
complicated mixture of non-thermal and thermal emission in the PSPC
means that a definitive result on the shape of the non-thermal X-ray
spectrum awaits X-ray data of higher spatial and spectral resolution
such as those which could be obtained with AXAF.

Using a computer code developed for the work of
\markcite{hard-hot98}Hardcastle, Birkinshaw \& Worrall (1998a), we
model the nuclear source as a spherical emitting region in which the
magnetic field and electron energy densities are in equipartition.
Assuming negligible bulk relativistic motion, we can place a lower
limit to the radius of the emitting region of 0.2~pc; for smaller
sizes the synchrotron self-absorption cut-off frequency would be above
2~GHz.  For a radius of 0.2~pc the equipartition magnetic field,
$B_{\rm eq}$, is $\approx 0.07$~Gauss, and the single-component
electron spectrum must span at least the range $E_{\rm min} \sim 5
\times 10^7$ to $E_{\rm max} \sim 10^{12}$ eV to give radiation at
$\sim 2$~GHz and $\sim 1$~keV.  The self-Compton contribution to the
X-ray emission is negligible, at $\lesssim 0.1\%$.  For larger
objects, $B_{\rm eq}$ scales as $(0.2 {\rm pc}/r)^{6/7}$ and $E_{\rm
min, max}$ scale as $(r/0.2 {\rm pc})^{3/7}$.

We argued in section~\ref{agnradx} that, since J2310-437's core is
relatively weak compared with its extended (isotropic) radio emission,
relativistic boosting along the line of sight is likely to be small
compared with most BL~Lac objects.  Relativistic boosting is
quantified by the bulk relativistic Doppler factor, $\delta$, which is
given by $1 / \gamma (1 - \beta \cos\theta)$, where $\gamma$ is the
Lorentz factor, $\beta$ is the velocity as a fraction of the speed of
light, and $\theta$ is the angle to the line of sight. 
For a steady jet of optically-thin spectral index $\alpha_s$ and jet to
counter-jet ratio $R$, $\beta \cos\theta = [(R^{1 \over 2 + \alpha_s} -
1)/(R^{1 \over 2 + \alpha_s} + 1)]$ (\markcite{sch79}Scheuer \&
Readhead~1979; \markcite{begel84}Begelman, Blandford \& Rees~1984).
Unified models of BL Lacs and radio galaxies with low
core to isotropic radio ratios suggest $\theta \gtrsim 30^\circ$
(\markcite{pad90}Padovani \& Urry 1990).  Our measurements of $R$
(section~\ref{overallradio}) then suggest that $\delta < 2$.

\begin{figure*}
\plotone{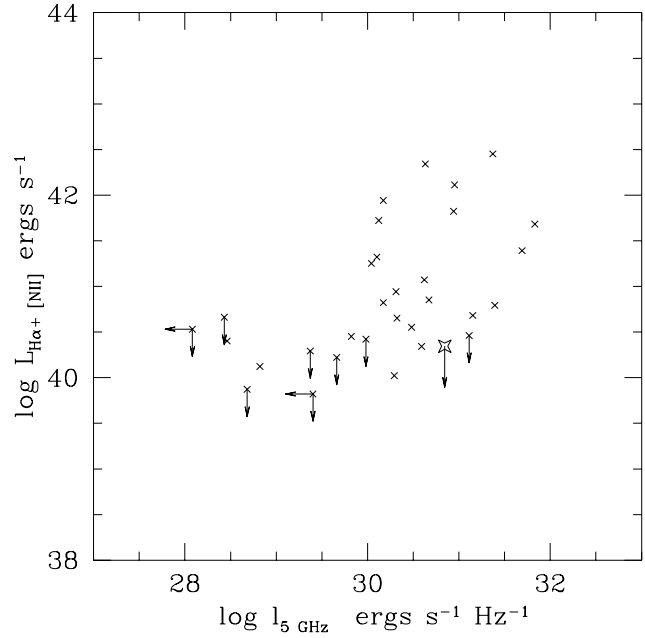}
\caption{H$\alpha$ + [N II] luminosity versus 5~GHz core radio power
for the B2-selected sample of low-power radio galaxies (crosses) from
Morganti et al.~(1992; adjusted to $H_o = 50$~km s$^{-1}$ Mpc$^{-1}$
used in this paper) and J2310-437 (star).
The radio to X-ray spectral index of 0.61 for J2310-437 is flatter
than
for the B2 radio galaxies (average value roughly 0.85) and so the
factor of $\sim 70$ times larger X-ray luminosity with respect to
a radio galaxy of the same radio core strength makes it surprising
that J2310-437 does not have lines of a detectable strength.
\label{lines}}
\end{figure*}

If the AGN in J2310-437 is BL~Lac-like, with a
relatively large orientation to the line of sight, we have to consider
why the X-ray emission is so bright (compared with an FR1~galaxy
nucleus), and why there is no evidence of optical continuum or line
emission.  Ulrich (1989) has shown that a sample of radio galaxies
selected from the B2 radio survey is particularly well matched to
BL~Lac objects in extended (isotropic) radio emission and galaxy
magnitude.  In Figure~\ref{lines} we compare our upper limit to the
emission-line luminosity of J2310-437 with measurements for B2
galaxies from Morganti, Ulrich \& Tadhunter~(1992). The plot shows
core radio power on the abscissa since a correlation with line
luminosity is claimed for larger samples of FR1 radio
galaxies (Baum, Zirbel \& O'Dea~1995).  We find that radio galaxies
with the same core radio power as J2310-437 generally have lines
detectable at levels higher than the upper limit for J2310-437.  If we
assume that the line emission in FR1 radio galaxies is due to
photoionization (but see Baum et al.~1995), then we might expect
J2310-437's line emission to be enhanced relative to galaxies of the
same radio-core power since its ratio of X-ray to radio strength is
higher ($\alpha_{\rm rx} = 0.61$) than for a typical FR1 radio galaxy
($\alpha_{\rm rx} = 0.85$). Perhaps this indicates that line-emitting
clouds in J2310-437 are indeed weak or absent. Baum et al.~(1995)
argue that the FR1/BL~Lac population is deficient in isotropic
radiant optical/UV energy with respect to higher-power radio
galaxies, possibly due to a lower accretion rate. 
J2310-437 may show an extreme ratio of bulk kinetic energy
in the jet to radiant isotropic energy close to the central AGN.

\begin{figure*}
\plotone{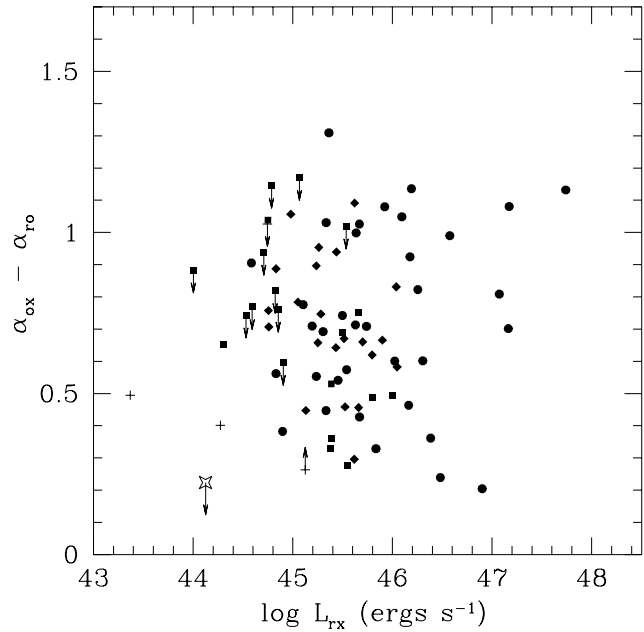}
\caption{Size of break in spectral index between 
$\alpha_{\rm ro}$ and $\alpha_{\rm ox}$
plotted against the logarithm of the sum of the 
two-point radio (5~GHz) to optical (4400~\AA) and
optical to X-ray (1~keV) luminosities for objects
from Figure~\ref{aoxplot} with measured
redshifts. J2310-437 (star) is at the low end of the XBL luminosity
range and shows little or no spectral curvature.
\label{breaklum}}
\end{figure*}

The possibility that J2310-437 exhibits a single-component synchrotron
spectrum from the radio to the X-ray distinguishes it from BL Lac
objects which show strong curvature and relatively more optical
emission.  Landau et al.~(1986) found the radio to X-ray spectrum of a
BL~Lac object was well described by a parabola. A remarkable
correlation between the half-width of the parabola and bolometric
luminosity of the source (Jones, Rudnick \& Landau 1986), in the sense
of less luminous sources displaying an overall flatter spectrum, led
these authors to suggest that the underlying physical quantities
determining the spectra from such sources are limited in number.
Worrall (1989) showed the X-ray emission in radio-selected BL~Lac
objects was consistent with an extrapolation of the parabola, but that
a spectral discrepancy hinted at a possible admixture of flat-spectrum
Compton emission.  In Figure~\ref{breaklum} we compare J2310-437 with
the objects from Figure~\ref{aoxplot} in plotting the size of the
break in spectral index between $\alpha_{\rm ro}$ and $\alpha_{\rm
ox}$ against the logarithm of the sum of the two-point radio (5~GHz)
to optical (4400~\AA) and optical to X-ray (1~keV) luminosity.  The
abscissa is a measure of the bolometric luminosity whereas the
ordinate is a measure of the degree of curvature in the spectrum.  A
correlation of greater curvature with larger luminosity is not
apparent in this figure, but the extreme position of J2310-437
compared with other XBLs is again obvious.

The work of Landau~et al.~(1986) has been extended by Sambruna et
al. (1996), who fit parabolas to to the spectra of BL~Lac objects and
core-dominated quasars and argue that lower-power sources have higher
peak frequencies in $\nu l_\nu$.  In J2310-437 we have found that the
break frequency may be within or above the X-ray band.  Since the lack
of line emission in J2310-437 may indicate an overall shortage of
optical and UV photons external to the jet in the central regions of
the source, the source's properties are consistent with the
intensity of external photons governing the electron spectral break
through Compton cooling: a higher external photon density produces a
lower-energy electron spectral break (Ghisselini 1997).  In J2310-437
the external photon density may be sufficiently low as to allow an
unbroken synchrotron spectrum from the radio to the X-ray.

We have interpreted the AGN in J2310-437 as a jet in which the
electron population produces X-ray synchrotron radiation more
effectively than in most BL~Lac objects and FR1 radio galaxies.  It is
important to ascertain how common or rare such sources are, as
discussed in section~\ref{agnopt}.  In particular, the discovery of
X-ray galaxies/AGN with optical continuum deficits, no emission lines,
but no core radio emission (i.e., radio-quiet versions of J2310-437) may
point to a completely different interpretation.  Such a population may
have been found by Griffiths et al.~(1995), in their so-called
`passive X-ray galaxies', although here the authors favor attributing
the strong X-ray emission to a hot coronal halo, possibly in the early
stages of a cooling flow, which is an unlikely explanation
for J2310-437.

\section{Conclusions}

The ROSAT HRI X-ray data and ATCA radio imaging and polarization
measurements confirm the presence of an AGN in the elliptical galaxy
J2310-437. The radio properties are reminiscent of an FR1 radio
galaxy, even in the sense that the cluster-scale X-ray emitting medium
around J2310-437 is overpressured with respect to the minimum pressure
in the radio structures.  There may be a cooling flow associated with
the cluster gas, but the central X-ray emission from this would
provide only a small contribution to the compact X-ray source.

The radio data, including the strength of the core relative to the
extended (isotropic) emission, suggest an ordinary FR1 radio galaxy
with its jet at $\gtrsim 30^\circ$ to the line of sight.  However, the
relative strength of the X-ray emission suggests we are observing a
low-power BL~Lac-type nucleus, and J2310-437's radio and X-ray
properties are not entirely without precedent for an X-ray selected
BL~Lac object.  What then makes the source extreme is a lack of the
expected level of optical continuum.  Whereas the multiwavelength
spectra of BL Lac objects fit parabolas between the radio and X-ray,
in J2310-437 a single power law connecting the radio and X-ray runs a
factor of about 2 below the optical upper limit, and it is possible to
interpret all of the emission as arising from a single region of
synchrotron-emitting plasma. J2310-437 is the most extreme such source
known.

Low-power sources are normally the most deficient in central isotropic
radiant optical/UV energy with respect to jet power, a feature which
Baum et al.~(1995) attribute to low accretion rate.  Moreover, it is
argued that the lowest-power sources have the highest (jet) peak
frequencies in $\nu l_\nu$ (Sambruna et al.~1996).  J2310-437 may be
interpreted as an extreme low-power source in which a single
synchrotron component extends to, and peaks in, the X-ray band, at
which point the PSPC data suggest spectral steepening may occur.
J2310-437's low isotropic optical/UV radiation is consistent with the
idea that the intensity of external photons governs the electron
spectral break through Compton cooling (Ghisselini~1997): in this
source the external photon density would be too low to produce a
spectral break below the X-ray. Such an explanation for the AGN in
J2310-437 would be challenged by the discovery of `radio-quiet'
versions of the source.

\acknowledgments

We thank G\"unther Hasinger for discussions concerning ROSAT's aspect
determination errors, and John Silverman and Dan Harris for providing
preliminary versions of IRAF scripts being developed to help correct
for such errors.  Martin Hardcastle provided software used to
calculate physical parameters for the radio emission.  RAR
acknowledges the excellent observer support provided by the staff at
CTIO.  DMW and MB thank the following `duty officers' and `friends'
for help before and during the ATCA observing run: Vince McIntyre,
John Reynolds, Tasso Tzioumis and Bob Sault.  The Australia Telescope
is funded by the Commonwealth of Australia for operation as a National
Facility managed by CSIRO. We acknowledge support from NASA grants
NAG~5-3205 and NAG~5-1882.  RAR's CTIO observing was supported by
travel grant NSF AST 9315074, and travel support for DMW and MB to
observe with the ATCA was provided by the PPARC.

\appendix

\section{Effect of ROSAT  Aspect Correction Errors on the J2310-437
HRI data} \label{appendix}

\markcite{david96}David et al.~(1996) illustrate the effects of
residual errors in the ROSAT aspect solution by showing HRI images of
6 stellar sources measured at small off-axis angles.  They point out
that the surface brightness at radii between 5 and 8 arcsec from the
center of the images can exhibit some asymmetry; the feature is
randomly oriented and cannot be predicted through knowledge of the
roll angle.  In particular, a source observed at different epochs,
with different roll angles, may be affected differently and
unpredictably. 

The problem responsible for this behavior has been described by
G. Hasinger in private discussions during 1997. It is believed that
there are gain changes between neighboring 1 arcmin$^2$ pixels in the
aspect camera, with possibly as many as 20-30\% of the pixels
returning inaccurate centroids.  The full aspect-camera data are not
returned in the telemetry stream, only the computed centroids. Problematic
pixels may then be used for aspect corrections for some phases of the
402~sec spacecraft wobble which, of amplitude about $\pm 3$ arcmin, is
employed when the HRI is in focus in order to average out any changes
in quantum efficiency over the detector.  Depending on the roll angle
and the positions of aspect stars, some observations are entirely free
from errors, although which observations are unaffected cannot be
predicted in advance, nor after the fact (except based on the quality
of the X-ray images).

Since the relative wobble phase is given by the spacecraft clock,
which is rarely reset, Hasinger's suggested procedure for improving the
aspect errors in a given data set is to fold the data on the wobble
period and divide into some number of phase bins.  A centroid should
be found for data in the resulting set of time intervals for each
phase bin in turn.  A shift should then be applied so that these centroids
are aligned.  Only data for the same roll angle (which changes in
increments of multiples of 1 degree) should be analyzed together.  The
amplitude of the wobble relative to the size of pixels in the aspect
camera implies that at least 10 phase bins should be used.
\markcite{morse94}Morse (1994) showed empirically that for a bright
stellar source the spread of counts sharpened significantly when
sub-images defined using detector coordinates as an indicator of
satellite wobble-position were selected and stacked to a common
centroid.  Morse suggests that for an HRI target of 0.35 cts/s (cf
0.1 cts/s for J2310-437), 20 to 30 bins should provide a reasonable
trade-off between fineness of phase grid and adequate counts for
determination of good centroids.

The 1996 observations of J2310-437 stretched over more than a month
and spanned 4 roll angles in a 20 degree band; a severe disadvantage.
Had there been, within about 6 arcmin of J2310-437, a bright X-ray
source identified with a point-like emitter, we could have used the
radial distribution of its counts as representative of the PRF for the
observation; unfortunately no such source is present.  So, assisted by
preliminary versions of IRAF scripts being written by John Silverman
and Dan Harris for distribution by the US-RSDC/SAO, we attempted the
`dewobble' procedure described above on the 1996 data of J2310-437.
Rather than analyze the data-set as one, we broke it into 5 time
intervals, each of constant nominal roll angle.  For the sake of
centroid accuracy it then seemed unreasonable to divide the data into
more than 10 phase bins.  We smoothed and contoured each of the
separate 50 images before re-stacking.  In only about half the cases
did this produce an image which was more radially symmetric than the
composite, and other cases were significantly more elongated.  This
suggests that 10 phase bins are too few, but the relative weakness of
the source coupled with the several changes in roll angle makes it a
poor candidate for finer binning.  Thus we are not able to correct for
the spread in the 1996 data introduced by residual aspect errors.

\begin{figure*}
\epsscale{.9}
\plotone{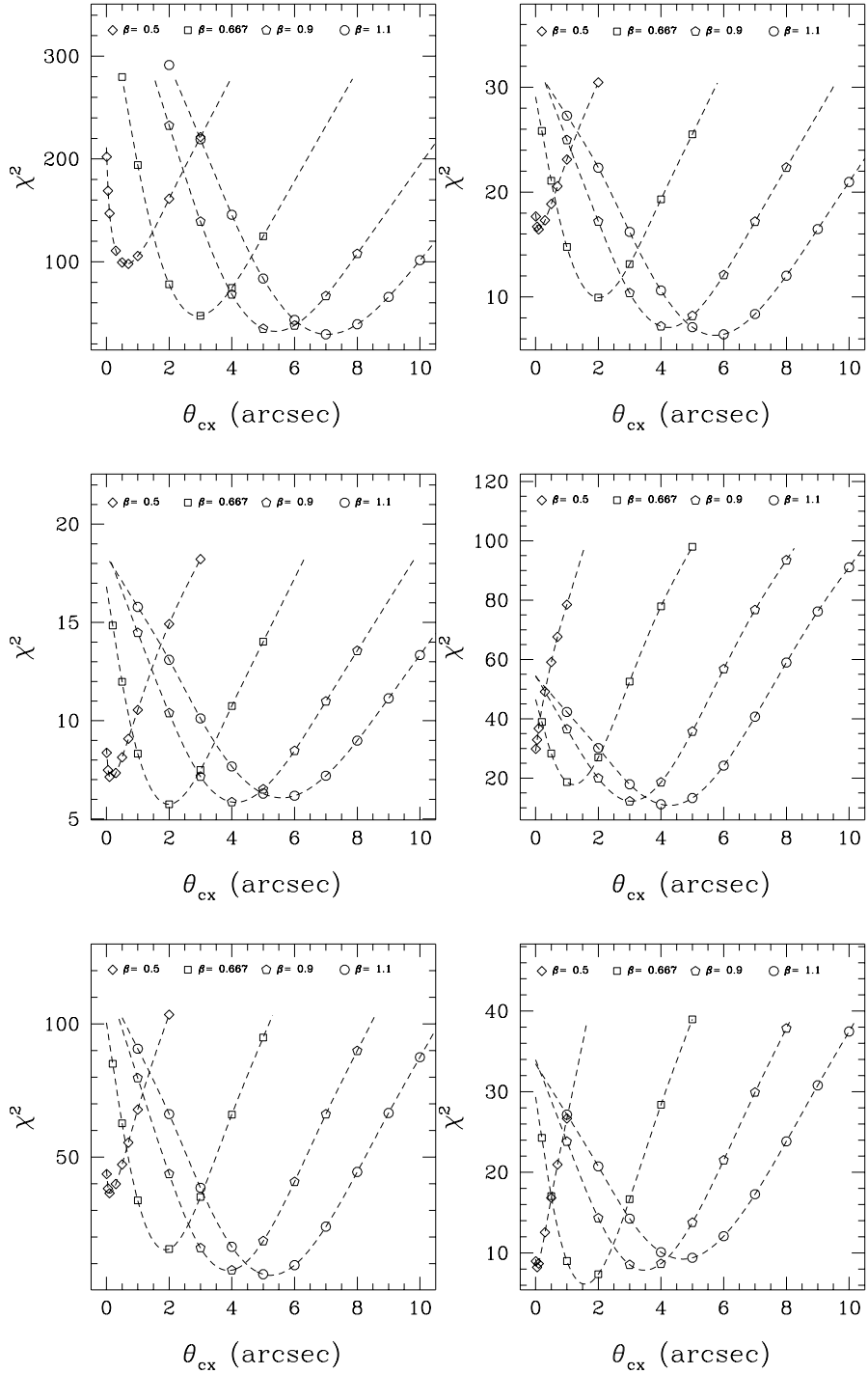}
\caption{Results of fitting a beta model convolved with the nominal
PRF to six subsections of the 1996 HRI data, each of duration no more
than 3 days. Each fit has 14 degrees of freedom.  The subsections of
data are not of equal length, and the net counts in the profiles,
clockwise starting with the top-left figure, are 958, 118, 295, 200,
418, and 90.
Since the 1995 HRI data indicate that the source in
reality should be unresolved to the HRI, these fits quantify
aspect smearing in terms of $\beta$-model parameter values.
\label{hri2chi}}
\end{figure*}

J2310-437 does, however, present a particularly interesting case in
that the 1995 data fit the nominal PRF so well.  This fact, coupled
with the similar flux measured in 1996, leads us to believe that
attitude smearing must be the cause of elongation in the 1996 data.
It is then useful to characterize the amount of smearing by fitting a
beta model convolved with the nominal PRF to the 1996 data; the
parameters found for the $\beta$ model are then of a size which we
would not necessarily believe as real if determined for some other
data set.  We have broken the data into 6 time intervals, where each
spans no more than 3 days.  For each we have extracted a radial
profile, using the method described for the 1995 data in the main body
of the text.  The net counts range between about 100 and 1000 counts,
depending on segment.  $\chi^2$ vs core radius is plotted in
Figure~\ref{hri2chi} for four different values of $\beta$.  The plots
indicate, for example, that a data set which fits a $\beta$ model of
$\beta = 2/3$ and core radius $\lesssim 3$~arcsec, or $\beta = 0.9$
and core radius $\lesssim 5$~arcsec, are within the regime where
aspect smearing could be responsible for the results.

\clearpage

\end{document}